\documentclass[11pt]{article}
\usepackage{comment}
\usepackage[margin=2.5cm]{geometry}
\usepackage{amsmath,amssymb,extarrows,mathtools,graphicx,subfigure,setspace,bbold,physics}
\usepackage{cite}
\usepackage{slashed}
\usepackage[dvipsnames]{xcolor}
\usepackage{hyperref}
\hypersetup{colorlinks=true, linkcolor=blue, citecolor=magenta, linktoc=page}
\usepackage{tikz}
\usepackage{arydshln,leftidx,mathtools}
\usetikzlibrary{decorations.pathreplacing}
\numberwithin{equation}{section}
\newcommand{\be}{\begin{equation}}
\newcommand{\bea}{\begin{eqnarray}}
\newcommand{\eea}{\end{eqnarray}}
\newcommand{\ba}{\begin{align}}
\newcommand{\ea}{\end{align}}
\newcommand{\ee}{\end{equation}}

\usepackage{chngcntr}
\counterwithout*{figure}{section}
\begin{document}

\begin{titlepage}
\thispagestyle{empty}

\begin{flushright}
IPM/P-2020/014\\
\end{flushright}

\vspace{.4cm}
\begin{center}
\noindent{\Large \textbf{Evolution of Entanglement Wedge Cross Section Following a Global Quench}}\\

\vspace*{15mm}
\vspace*{1mm}
{Komeil Babaei Velni${}^{\ast, \dagger, \natural}$, M. Reza Mohammadi Mozaffar${}^{\ast, \dagger}$ and M. H. Vahidinia${}^{\ddagger, \dagger}$}

 \vspace*{1cm}

{\it  ${}^\ast$ Department of Physics, University of Guilan,
P.O. Box 41335-1914, Rasht, Iran\\
${}^\dagger$ School of Physics,
Institute for Research in Fundamental Sciences (IPM),
P.O. Box 19395-5531, Tehran, Iran\\
${}^\ddagger$ Department of Physics, Institute for Advanced Studies in Basic Sciences (IASBS),\\ P.O. Box  45137-66731,  Zanjan, Iran\\
${}^\natural$ School of Particles and Accelerators, Institute for Research in Fundamental Sciences (IPM), P.O. Box 19395-5531, Tehran, Iran
}

 \vspace*{0.5cm}
{E-mails: {\tt babaeivelni@guilan.ac.ir, mmohammadi@guilan.ac.ir, vahidinia@iasbs.ac.ir}}%

\vspace*{1cm}
\end{center}

\begin{abstract}

We study the evolution of entanglement wedge cross section (EWCS) in the Vaidya geometry describing a thin shell of null matter collapsing into the AdS vacuum to form a black brane. In the holographic context, it is proposed that this quantity is dual to different information measures including entanglement of purification, reflected entropy, odd entropy and logarithmic negativity. In $2+1$ dimensions, we present a combination of numerical and analytic results on the evolution and scaling of EWCS for strip shaped boundary subregions after a thermal quench. In the limit of large subregions, we find that the time evolution of EWCS is characterized by three different scaling regimes: an early time quadratic growth, an intermediate linear growth and a late time saturation. Further, in $3+1$ dimensions, we examine the scaling behavior by considering thermal and electromagnetic quenches. In the case of a thermal quench, our numerical analysis supply results similar to observations made for the lower dimension. On the other hand, for electromagnetic quenches at zero temperature, an interesting feature is a departure from the linear behavior of the evolution to logarithmic growth.
\end{abstract}

\end{titlepage}

\newpage

\tableofcontents
\noindent
\hrulefill

\onehalfspacing

\section{Introduction}\label{intro}
Non-equilibrium dynamics and thermalization of a strongly coupled system is a long-standing problem in many areas of physics. In the holographic context, equilibration from a highly excited initial state is expected to be dual to black hole formation under a gravitational collapse. So in this scenario, issues about the black hole physics are tightly connected to the physics of thermalization in the dual strongly coupled system \cite{hep-th/9912209} (see \cite{1810.02367} for review).

A simple setting that shows the general features of equilibration in a far-from-equilibrium system is a global quench. In this setup, one considers the creation of a homogeneous and isotropic highly excited state from the vacuum state by an abrupt change in the Hamiltonian of a closed quantum system. It is expected that this excited state evolves towards the equilibrium and shows some aspects of a thermalization process \cite{0003193, cond-mat/0503393} (refer to \cite{0905.4013,1007.5331,1603.02889} for review). Remarkably, the holographic dual of this dynamics is simply described by the Vaidya geometry that shows the collapse of a thin shell of null matter and black hole formation which is an exact solution to Einstein's theory of gravity \cite{hep-th/9912209,hep-th/0610041,0705.0016,0808.0910}. 
It should be noted that the Vaidya geometry is not the only possible holographic model describing the quantum quench, in particular, it can be modeled namely by the time evolution of black hole interior \cite{Hartman:2013qma}. 

One may study the dynamics of a globally quenched system by evaluating the correlation between the subsystems of a given system  \cite{cond-mat/0601225}. Among other things, it is known that the entanglement entropy (EE) is a useful probe to capture this dynamics \cite{cond-mat/0503393}. EE measures the quantum correlations between a subsystem $A$ and its complement $A^{c}$ and is defined as von Neuman entropy of a reduced density matrix $\rho_{A}=\tr_{A^{c}}(\rho)$ as
\begin{equation}
S_A=- \tr(\rho_A \ln \rho_A).
\end{equation}
As the non-equilibrium system evolves towards the equilibrium, the  EE grows with time linearly 
and saturates at the equilibrium value which is equal to thermal entropy. This behavior of EE growth has a simple description in terms of propagating entangled pairs of quasi-particles \cite{cond-mat/0503393,0905.4013}. 

In quantum field theories with holographic dual, there is a very interesting prescription for computing EE. According to Ryu and Takayanagi (RT) seminal proposal \cite{hep-th/0603001}, EE corresponding to a spatial subregion $A$ in the CFT is  given by the area of a  codimension-2 minimal surface $\Gamma_{A}$ 
\begin{equation}\label{RT}
S_{A}=\frac{{\rm area}(\Gamma_A)}{4G_N},
\end{equation}
where the bulk minimal RT surface $\Gamma_{A}$ is homologous to the subregion $A$ such that its boundary anchored to the boundary of $A$ ($\partial A=\partial \Gamma_A$). The authors of \cite{0705.0016} generalized this prescription to the time-dependent backgrounds by assuming  $\Gamma_{A}$ is an extremal surface (HRT surface) subject to the same boundary condition. It is worthwhile to mention that both proposals have been derived in the context of AdS/CFT in \cite{1304.4926,1607.07506}.
	
Studying holographic entanglement entropy for an asymptotically AdS Vaidya geometry nicely captures the time evolution of EE in the dual CFT \cite{Lopez,Albash:2010mv,Balasubramanian:2010ce,Balasubramanian:2011ur,Liu:2013iza,Liu:2013qca}. In general, where the characteristic size of the boundary entangling region is large compared to inverse
temperature, EE shows a quadratic growth in the pre-local-equilibration and it follows by a linear growth regime in the past-local-equilibration.  After that and before saturation, the system evolves to memory loss regimes in which the EE forgets the size and shape of the entangling region. This behavior may suggest a simple geometric interpretation for the growth of entanglement based on the propagation of an entanglement wave with a sharp wavefront inward the entangling region from the entangling boundary. This model is dubbed as an “entanglement tsunami” \cite{Liu:2013iza,Liu:2013qca} (see also \cite{Casini:2015zua,Mezei:2016zxg,Mezei:2018jco,1912.11024,Alishahiha:2014cwa,Fonda:2014ula,Leichenauer:2015xra} for related studies).
 However, as authors of \cite{Kundu:2016cgh} explain, the tsunami picture only works for large entangling regions and breaks down for small regions. In this condition,  the saturation time is less than the equilibration time scale and it implies that the dynamics of the quenched system is governed by a different mechanism.

As mentioned, one way to study the equilibration process is the evaluation of the correlation between subsystems. For a pure state, EE measures the correlations between a subsystem and its complement. However, to analyze the correlation of two disjoint intervals, $A$ and $B$, EE is not a convenient quantity.  This is because EE is a measure of the (quantum) correlation when the total system is pure (subsystem and its complement) while for two disjoint regions $A$ and $B$, $\rho_{A \cup B}$ is not pure. In this case, one useful quantity is mutual information (MI) that measures the total correlation between two subsystems $A$ and $B$  
	\begin{equation}
	I(A, B)=S_{A}+S_{B}-S_{A \cup B}.
	\end{equation}
As MI defined in terms of EE, one may use the HEE proposal to study MI and its time evolution in the holographic setups \cite{1110.0488}. Further generalizations of MI to systems consisting of more (disjoint) subsystems, e.g. tripartite and $n$-partite information is studied in several directions in \cite{Allais:2011ys,Alishahiha:2014jxa}.
	
In the framework of holographic theories, EE and MI are related to (H)RT surfaces. Recently, a generalization of (H)RT surface which is called the entanglement wedge cross section (EWCS) has attracted a lot of attention. This geometrical quantity is defined as \cite{1708.09393,1709.07424}
\begin{equation}\label{EW}
E_W(A, B)=\frac{\text{area}(\Sigma_{AB}^{\min})}{4G_N},
\end{equation}
where $\Sigma_{AB}^{\min}$ is the minimal cross-sectional area of the entanglement wedge \cite{1211.3494,1204.1330,1408.6300} corresponding to the boundary region $A\cup B$.  
As shown in \cite{1708.09393}, the EWCS
is subject to the following inequalities\footnote{For a complete set of inequalities see \cite{1709.07424}.}
\bea\label{inequality}
\frac{I(A, B)}{2}\leq E_W(A, B)\leq {\rm min}\left(S_A, S_B\right),\label{ineq1}\\
E_W(A, B\cup C)\geq \frac{I(A, B)}{2}+\frac{I(A, C)}{2}\label{ineq2}\\
E_W(A, B\cup C)\geq E_W(A, B)\label{ineq3}.
\eea
One may argue that EWCS takes into account the correlation between the boundary subsystems $A$ and $B$ even for a mixed state \cite{1708.09393,1709.07424,1907.12555}. Therefore, it should be useful to probe the equilibration in a holographic system. The main goal of this paper is to study the dynamics of $E_W$ in the Vaidya background as a dual description of a global quench in a CFT.   

There are several proposals for the CFT counterpart of $E_W$. Initially, it was introduced as a possible dual of the entanglement of purification. This conjecture was based on some information theoretic properties and intuition from holographic tensor networks \cite{1708.09393,1709.07424}. However, it is turned out that several other correlation measures such as reflected entropy \cite{1905.00577},  logarithmic negativity \cite{1808.00446,1907.07824} and odd entropy \cite{1809.09109} also relate to $E_W$.  Interestingly, all of these measures  are useful for analyzing the correlation between $A$ and $B$ where $\rho_{A \cup B}$ is a mixed state. See \cite{Espindola:2018ozt,1805.02625,Agon:2018lwq,1812.05268,1902.02243,1902.02369,1902.04654,1904.09582,2001.05501} for related progress. In the following, we review each one of them briefly.

The entanglement of purification is a measure of classical and quantum correlations between two subsystems \cite{Terhal:2002}. 
To define entanglement of purification  let us assume that  $\rho_{AB}$ is (a mixed) density matrix for $A\cup B$  in  the total Hilbert space  $\mathcal{H}=\mathcal{H}_{A}\otimes\mathcal{H}_B$.  By adding some auxiliary degrees of freedom to $\mathcal{H}$, it is possible to construct a pure state $\dyad{\psi}$ such that $\rho_{AB}=\tr_{A'B'} (\dyad{\psi})$ and $\ket{\psi} \in \mathcal{H}_{AA'} \otimes \mathcal{H}_{BB'}$. Although this purification is not unique, one may consider a specific purification that minimizes the EE between $A$ and its auxiliary partner $A'$. Therefore,  the entanglement of purification  is defined as 
\begin{equation}
E_{P}(A, B)=\underset{ \;\rho_{AB}=\tr_{A'B'} (\dyad{\psi})}{\min} S_{AA'}.
\end{equation}
Clearly, the above definition reduces to EE when $\rho_{AB}$ is pure. Note that, this quantity should be minimized over all possible $\ket{\psi}$ so it is not an easy task to compute it in an arbitrary quantum theory. But under some assumptions one may investigate it in certain situations \cite{1709.07424,1803.10539}.  Moreover,  for holographic theories, it has been conjectured that entanglement of purification  is dual to the  area of entanglement wedge cross section $E_P=E_W$. In this case, minimization is restricted to states with holographic dual \cite{1708.09393,1709.07424}. 

More recently however, increasing attention has been paid to the reflected entopy as a new measure of the correlation between two disjoint regions. To define this measure, note that one can canonically purify the mixed state $\rho_{AB}=\sum_{i}p_{i} \ket{\rho}\bra{\rho}$ by doubling the Hilbert space $\mathcal{H}=\mathcal{H}_{A}\otimes\mathcal{H}_B$ such that
$\sqrt{\rho}=\sum_{i} \sqrt{p_i} \rho_{i}\otimes\rho_{i}$ be a pure state in $\mathcal{H}\otimes \mathcal{H'}$. Now the reflected entropy between $A$ and $B$ is defined as the entanglement entropy between $A$ and $A'$
\begin{equation}
S_{R}(A,B)\equiv- \tr(\rho_{AA'} \ln \rho_{AA'}), \quad \sqrt{\rho_{AA'}}=\tr_{BB'} \; \ket{\sqrt{\rho}}\bra{\sqrt{\rho}}.
\end{equation}
One may note that similar to the entanglement of purification, the reflected entropy also reduces to EE for pure states. Interestingly, it is possible to calculate this mesuare by using the replica method and some holographic argument shows that $S_R=2E_W$ \cite{1905.00577}. 

The logarithmic negativity is another quantity that captures the correlation between $A$ and $B$ but unlike mutual information, the entanglement of purification and reflected entropy, it is monotonic under local operations and classical communication (LOCC) and so is appropriate to capture quantum correlations for mixed states. It is defined as $\mathcal{E}(A, B)=\log \tr\qty(\rho_{AB}^{T_{B}})$, where $\rho_{AB}^{T_{B}}$ represents the partial transpose of $\rho_{AB}$ with respect to $B$ \cite{quant-ph/0505071}. Initially, based on the holographic quantum error-correcting code, authors of \cite{1808.00446} conjectured relation between the logarithmic negativity and the area of a brane with tension in the entanglement wedge. For the vacuum state and ball-shaped subregions, this reduces to 
$\mathcal{E}=\chi_{d} E_W$ where $\chi_d$ is a dimensional dependent constant.
Remarkably,  this relation has been derived by noting the connection between logarithmic negativity and R\'enyi reflected entropy and using the holographic prescription for computing R\'enyi entropy \cite{1907.07824}.

Finally, it is also worthwhile mentioned that there is also a new measure of correlations for mixed states which is called odd entropy \cite{1809.09109}
\begin{equation}
S_o(A, B)\equiv  \lim_{n_o\to1} \frac{1}{1-n_o}\qty[\tr\qty(\rho_{AB}^{T_B})^{n_o}-1].
\end{equation}
Based on holographic replica trick, odd entropy is related to the $E_W(A,B)$ and HEE between $A$ and $B$ as follows
\begin{equation}\label{odd}
S_{o}(A,B)=S(A,B)+E_W(A, B).
\end{equation}
One may note that $E_W(A,B)$ vanishes either for product state $\rho_{AB}=\rho_{A}\otimes\rho_{B}$ or pure state $\rho_{AB}= \ket{\psi}\bra{\psi}$, so $S_o(A,B)$ reduces to the von Neumann entropy in the former and to the EE in the latter. Refer to \cite{2004.04163} for recent study on this topic.

As mentioned, the area of EWCS should be a good geometrical quantity to capture correlations of mixed states in the dual quantum field theory and so it should be a useful tool to analyze the equilibration process after a quantum quench. Some authors have investigated aspects of this scenario mainly in two-dimensional CFTs. The author of \cite{Moosa:2020vcs} has investigated the time evolution of reflected entropy and its holographic dual after a global quench in the context of the thermal double model.
In a related study, the authors of \cite{2001.05501} have studied the dynamics of logarithmic negativity, odd entropy and reflected entropy as well as their holographic counterpart EWCS via AdS/BCFT after local and (in)homogenous global quenches. Furthermore, the time evolution of reflected and odd entropies under local quenches has been analyzed in \cite{1907.06646,1909.06790} where local quench is modeled by a falling particle in the holographic bulk theory. Also, the time evolution of EWCS in a two-sided black hole and Vaidya geometry has been studied in \cite{1810.00420}. 

In the current article, we aim to provide a detailed analysis of the time evolution of EWCS in various time-dependent geometries using holographic prescription. In particular, we are interested in various scaling regimes in the  EWCS dynamics during the thermalization process. For this purpose, we investigate EWCS for a strip-shaped region in the Vaidya geometry describing the collapse of a thin shell of null (charged) matter into the AdS vacuum to form a black brane as a holographic description of thermal (electromagnetic) quench.

The organization of the present paper is as follows. In section \ref{setup}, we give the general framework in which we are working, establishing our notation and the general form of the HEE and EWCS functionals both in static and time-dependent geometries. Section \ref{static} contains a brief summary about EWCS in static backgrounds which are dual to the initial and final equilibrium states. We review old results for AdS and AdS black brane geometries and also find new ones for the case of extremal black branes. In section \ref{3dim}, we investigate the time evolution of EWCS in $2+1$ dimensions, where we present both numerical and analytic results. Next, we study the higher dimensional cases by considering both thermal and electromagnetic quenches in section \ref{highdim}. We review our main results and discuss their physical
implications in section \ref{dis}, where we also present some future directions.

\section{Set-up}\label{setup}
We consider Einstein gravity coupled to a Maxwell field in (d+1)-dimensional asymptotically AdS spacetime. The action is
\begin{equation}
\label{polyaction}
I=\frac{1}{16 \pi G_N}\int d^{d+1}x \sqrt{-g}\left[R-2\Lambda-\frac{1}{4}F_{\mu\nu}F^{\mu\nu}\right],
\end{equation}
where $R$ is the Ricci scalar and $\Lambda=-\frac{d(d-1)}{2L^2}$ is the cosmological
constant, with $L$ being the AdS radius. The equations of
motion following from this action are solved by the geometry of a charged black brane
\begin{equation}\label{staticmetric}
ds^2=\frac{L^2}{r^2}\left(-f(r)dt^2+\frac{dr^2}{f(r)}+\sum_{i=1}^{d-1}dx_i\right),
\end{equation}
with
\begin{equation}\label{staticmetric1}
f(r)=1-m r^d+\frac{d-2}{d-1}q^2 r^{2d-2},\;\;\;A_t(r)=\mu\left(1-\left(\frac{r}{r_h}\right)^{d-2}\right),\;\;\;\mu=qr_h^{d-2},
\end{equation}
where $r_h$ denotes the horizon radius determined by the largest positive root of the blackening factor and $\mu$ corresponds to the chemical potential.\footnote{Without loss of generality we will from now on consider $L=1$.} This geometry is dual to a boundary theory at a finite density with the following expressions for energy, entropy and charge densities, respectively
\bea\label{ESC}
\mathcal{E}=\frac{d-1}{16\pi G_N}m,\;\;\;s=\frac{1}{4G_N}\left(\frac{1}{r_h}\right)^{d-1}, \;\;\;\rho=\frac{d-2}{4\pi G_N}\mu r_h^{d-2}.
\eea
Further, the Hawking temperature of the black brane is given by
\bea\label{T}
T=\frac{d}{4\pi r_h}\left(1-\frac{(d-2)^2}{d(d-1)}q^2r_h^{2(d-1)}\right).
\eea
Next, we will consider the extremal limit of the charged black branes where the temperature vanishes. It is straightforward to show that in this limit the blackening factor becomes
\bea\label{fext}
f_{\rm ext.}(r)=1-\frac{2d-2}{d-2} \left(\frac{r}{r_h}\right)^d+\frac{d}{d-2} \left(\frac{r}{r_h}\right)^{2d-2}.
\eea
Promoting the mass and charge in eq.\eqref{staticmetric} to time-dependent functions $m(v)$ and $q(v)$, the Vaidya solution is obtained where in the Eddington-Finkelstein coordinate is given by
\be
\label{vaidyametric}
ds^2=\frac{1}{r^2}\left[-f(r,v)dv^2-2dv dr +dx_{d-1}^2\right],\;\;\;\;f(r,v)=1-m(v) r^d+\frac{d-2}{d-1}q(v)^2 r^{2d-2}.
\ee
Here $v$ is a new coordinate defined by
\bea\label{vt}
dv=dt-\frac{dr}{f},
\eea
which coincides with the boundary time, i.e., $t$, at $r=0$.  This geometry describes an infalling null shell in an asymptotically AdS background. 

\begin{figure}
\begin{center}
\includegraphics[scale=0.7]{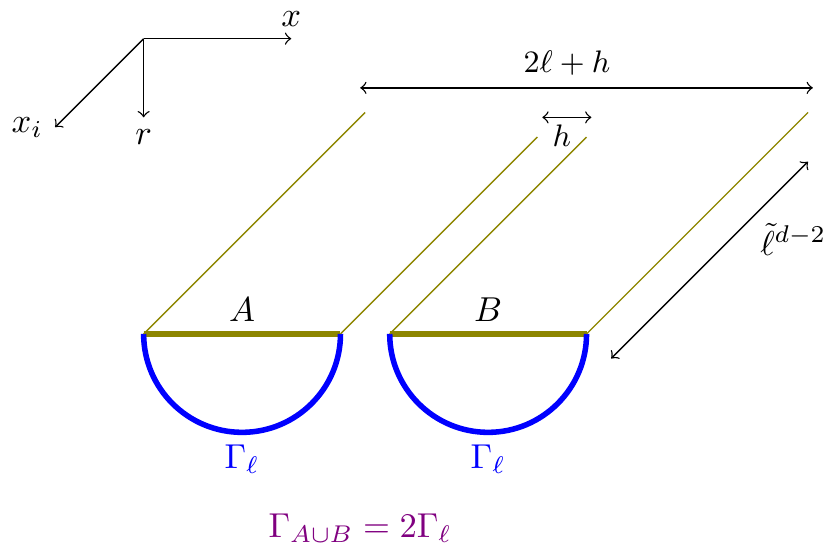}
\hspace*{0.2cm}
\includegraphics[scale=0.7]{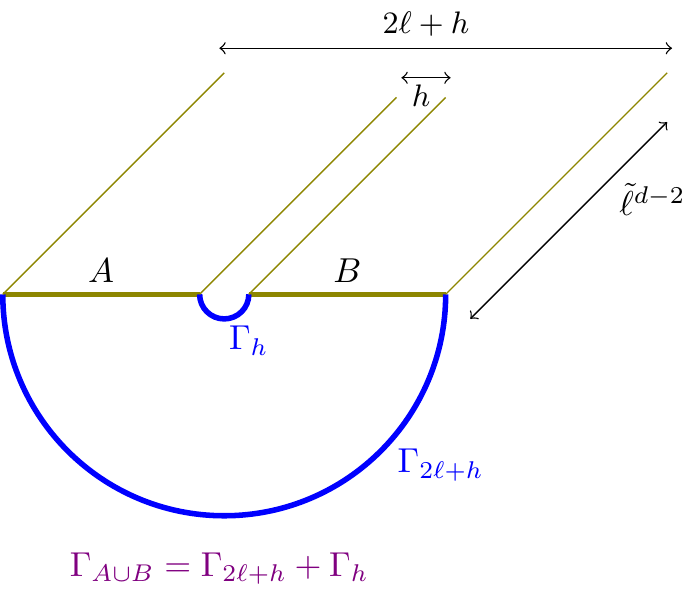}
\hspace*{0.2cm}
\includegraphics[scale=0.7]{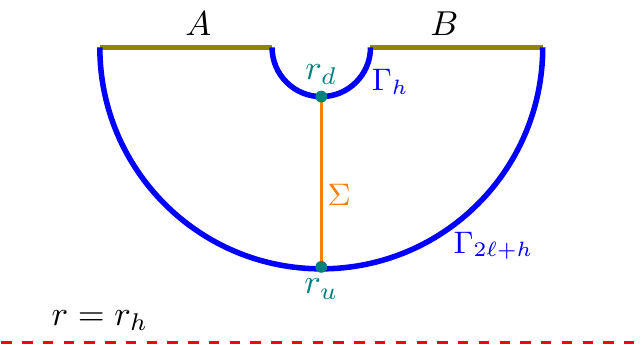}
\end{center}
\caption{Schematic minimal surfaces for computing $S_{A\cup B}$ in disconnected (left) and connected (middle) configurations. In the right panel, we show the EWCS ($\Sigma$ in orange). Here we only consider the connected configuration where the EWCS is non-zero.}
\label{fig:regions}
\end{figure}

In the next sections, we apply holographic prescription to find the time evolution of EWCS using eq. \eqref{EW} for configurations consisting of thin long strips. Figure \ref{fig:regions} shows the entangling regions that we consider for computing HEE and EWCS in the static geometry. When the entangling region in the boundary theory is a strip the corresponding domain is specified by 
\bea\label{stripregion}
-X/2\leq x_1\equiv x\leq X/2,\;\;\;\;\;\;\;\;\;0\leq x_i\leq \tilde{\ell},\;\;\;\;\;{\rm for}\;\;i=2,\cdots,d-1,
\eea
where $X$ and $\tilde{\ell}$ is the width and length of the strip, respectively. Note that in our set-up where we consider two different extremal hypersurfaces, i.e, $\Gamma_h$ and $\Gamma_{2\ell+h}$, $X$ is replaced with $h$ and $2\ell+h$, respectively. 
In $X\ll\tilde{\ell}$ limit, the translation invariance implies that the minimal hypersurface for computing HEE, i.e., $\Gamma_X$, is completely specified by $x(r)$. The HEE functional for static geometries eq.\eqref{staticmetric} then becomes
\bea\label{staticHEE}
S=\frac{{\tilde{\ell}}^{d-2}}{2G_N}\int_{0}^{r_t}\frac{dr}{r^{d-1}}\sqrt{\frac
{1}{f(r)}+x'(r)^2},
\eea
where $r_t$ is the turning point of the minimal hypersurface, i.e., tip of $\Gamma_X$. Of course, using the equation of motion, $r_t$ can be implicitly expressed in terms of $X$ as follows
\bea\label{staticel}
X=2\int_0^{r_t}\frac{dr}{\sqrt{\left(\left(\frac{r_t}{r}\right)^{2d-2}-1\right)f(r)}}.
\eea
On the other hand, due to the reflection symmetry the corresponding hypersurface for EWCS, i.e., $\Sigma$, lies entirely on $x=0$ slice.  In this case the EWCS functional becomes
\bea\label{staticEW}
E_W=\frac{{\tilde{\ell}}^{d-2}}{4G_N}\int_{r_t^{[h]}}^{r_t^{[2\ell+h]}} \frac{dr}{r^{d-1}\sqrt{f(r)}},
\eea
where $r_t^{[X]}$ is the turning point of the minimal hypersurface $\Gamma_X$ corresponding to a boundary region with width $X$. In the following, we will denote the $r_t^{[h]}$ and $r_t^{[2\ell+h]}$ as $r_d$ and $r_u$, respectively.
In \cite{1708.09393,1709.07424} it was shown that EWCS exhibits a discontinuous phase transition which is due to the competition between two different configurations for computing $S_{A\cup B}$. At small distances, i.e., $h\ll \ell$, a connected configuration $(\Gamma_{A\cup B}=\Gamma_{2\ell+h}+\Gamma_{h})$ has the minimal area, while for large separations the RT surface changes topology and the disconnected configuration $(\Gamma_{A\cup B}=2\Gamma_{\ell})$ is favored. In the latter case $\Sigma$ becomes empty and hence the EWCS vanishes (see figure \ref{fig:regions}). Indeed, this behavior is similar to the continuous phase transition of holographic mutual information (HMI) and the corresponding critical points are exactly the same. In order to have a nontrivial $\Sigma$ and nonvanishing $E_W$, we consider the small separation limit $h\ll \ell$ in the following.

Let us now turn to the time-dependent case where the geometry in the bulk is given by a Vaidya spacetime. Once again, the translation invariance implies that the extremal hypersurface is completely specified by $r(x)$ and $v(x)$. Using eq. \eqref{vaidyametric}, the HEE functional  can be written as
\bea\label{HEE}
S=\frac{{\tilde{\ell}}^{d-2}}{4G_N}\int_{-X/2}^{X/2} dx\; \frac{\mathcal{S}}{r^{d-1}},\;\;\;\;\;\mathcal{S}=\sqrt{1-2\dot{v}\dot{r}- \dot{v}^2 f(r,v)},
\eea
where $\dot{}\equiv \frac{d}{dx}$. Extremizing the above expression yields the equations of motion for $r(x)$ and $v(x)$, which read
\bea\label{EOMHEE}
\frac{\dot{v}^2}{2r^{d-1}\mathcal{S}}\frac{\partial f}{\partial v}=\frac{\partial }{\partial x}\left(\frac{\dot{r}+\dot{v}f}{r^{d-1}\mathcal{S}}\right),\;\;\;\;\frac{\dot{v}^2}{2r^{d-1}\mathcal{S}}\frac{\partial f}{\partial r}+\frac{(d-1)\mathcal{S}}{r^d}=\frac{\partial }{\partial x}\left(\frac{\dot{v}}{r^{d-1}\mathcal{S}}\right).
\eea
In this case, the corresponding boundary conditions for the extremal hypersurface are given as follows
\bea\label{bdycondition}
r(0)=r_t,\;\;\;\;\;v(0)=v_t,\;\;\;\;\;\dot{r}(0)=0,\;\;\;\;\;\dot{v}(0)=0,\;\;\;\;\;r(\frac{X}{2})=0,\;\;\;\;\;v(\frac{X}{2})=t,
\eea
where $(r_t,v_t)$ is the location of the turning point. On the other hand, the EWCS can be parametrized as $v=v(r)$ where due to the reflection symmetry the corresponding hypersurface, i.e., $\Sigma$ will be symmetric with respect to the midpoint $x=0$. In this situation, the EWCS functional becomes
\bea\label{funcEWCS}
E_W=\frac{{\tilde{\ell}}^{d-2}}{4G_N}\int dr\frac{\mathcal{F}}{r^{d-1}},\;\;\;\;\;\;\mathcal{F}=\sqrt{-2v'-f(r,v){v'}^2}.
\eea
The equation of motion obtained extremizing the above functional is
\bea\label{EOMEWCS}
\frac{\partial}{\partial r}\left(\frac{1+f(r,v) v'}{r^{d-1}\mathcal{F}}\right)=-\frac{\partial }{\partial v}\left(\frac{\mathcal{F}}{r^{d-1}}\right),
\eea
where the hypersurfaces of interest satisfy the following boundary conditions
\bea\label{bdyconditionEWCS}
v(r_d)=v_d,\;\;\;\;v(r_u)=v_u.
\eea
Now we are equipped with all we need to calculate the time dependence of HEE and EWCS using eqs. \eqref{HEE} and \eqref{funcEWCS}, respectively.
Unfortunately, it is not possible to find the time evolution of HEE and EWCS during the thermalization process analytically in general dimensions. In the following we will present the numerical results in the thin shell regime. Assuming this condition, the background that we consider for the thermal quench is given by eq. \eqref{vaidyametric} with 
\bea\label{mq}
m(v)=\frac{m}{2} \left(1+\tanh\left(\frac{v}{v_0}\right)\right),
\eea
where $v_0\ll 1$ is the parameter that controls the thickness of the null shell. Note that in this setup $v=0$ denotes the location of the null shell. 
We comment on $q(v)$ in the case of electromagnetic quench later.

\section{Preliminaries: EWCS for static backgrounds}\label{static}
Before examining the full time-dependence of $E_W$, we would like to study its asymptotic behaviors where the geometry is static. This study plays an important role in our analysis in the next sections because according to eqs. \eqref{vaidyametric} and \eqref{mq}, the early and late time geometries correspond to a pure AdS and a charged AdS black brane, respectively. So in the following we review the computation of $E_W$ in these backgrounds. Under these circumstances, the corresponding extremal hypersurface, i.e., $\Sigma$ lies entirely on a constant time slice inside the bulk. 
In the subsequent subsections, we present two specific examples in which we evaluate the behavior of $E_W$. We will consider AdS-Schwarzschild and Extremal AdS black brane geometries for which semi-analytic results can be obtained.

\subsection{AdS-Schwarzschild Black Brane}\label{sec:adsbb}
For the AdS-Schwarzschild black branes, the EWCS can be evaluated analytically in different scaling regimes. In this case, we consider $f(r)=1-\left(\frac{r}{r_h}\right)^d$ where $r_h=m^{-1/d}$ is the horizon radius. Evaluating eq. \eqref{staticEW} gives an exact result\cite{1810.00420}
\bea\label{ewBB}
E_W=\frac{\tilde{\ell}^{d-2}}{4(2-d)G_N}\left(\frac{\sqrt{f(r)}}{r^{d-2}}-\frac{d-4}{4}\frac{r^2}{r_h^{d}} \,{_2F_1}\left(\frac{1}{2},\frac{2}{d},\frac{d+2}{d},\frac{r^{d}}{r_h^{d}}\right)\right)\bigg|_{r_d}^{r_u}.
\eea
On the other hand, the relation between the position of the turning point $r_t$ and the strip width $X$ can be written as follows \cite{Fischler:2012ca}
\bea\label{ellrt}
X=2r_t\sum_{n=0}^{\infty}\frac{1}{1+nd}\frac{\Gamma\left(n+\frac{1}{2}\right)}{\Gamma\left(n+1\right)}\frac{\Gamma\left(\frac{d(n+1)}{2d-2}\right)}{\Gamma\left(\frac{dn+1}{2d-2}\right)}\left(\frac{r_t}{r_h}\right)^{nd},
\eea
where the infinite series converges for $r_t<r_h$. In principle, we can invert this formula to write eq. \eqref{ewBB} in terms of the boundary quantities, $h$, $2\ell+h$ and $T$. 
For the sake of simplicity, in the following we will focus on the low and high temperature behavior of EWCS.
As demonstrated in\cite{BabaeiVelni:2019pkw} considering low temperature with respect to the separation scale corresponds to $h\ll T^{-1}\ll \ell$. On the other hand, one might also regard the $h\ll \ell\ll T^{-1}$ case where we have low temperature with respect to both the subregion sizes and the separation between them. Further we note that, $h\ll T^{-1}\ll \ell$ limit corresponds to $r_d\ll r_h$ and $r_u\rightarrow r_h$, while for $h\ll \ell\ll T^{-1}$ we have $r_d, r_u\ll r_h$. Now using eqs. \eqref{ewBB} and \eqref{ellrt} one can find that the low and high temperature expansion of EWCS for $d>2$ is given by (see \cite{BabaeiVelni:2019pkw} for details)
\bea\label{ewlowhigh}
E_W\sim \Bigg\{ \begin{array}{rcl}
&E_W^{\rm vac.}-\alpha \frac{\tilde{\ell}^{d-2}}{G_N}\ell(\ell+h)T^{d}+\cdots,&\,\,\,h\ll \ell \ll T^{-1}\\
&\frac{\tilde{\ell}^{d-2}T^{d-2}}{4G_N}\left(-\frac{\beta}{(hT)^{d-2}}+\gamma+\lambda(hT)^2\right),
 &\,\,\,h\ll T^{-1}\ll \ell 
\end{array}.
\eea
where $\alpha$, $\beta$, $\gamma$ and $\lambda$ are some constants and $E_W^{\rm vac.}$ is the vacuum (pure AdS) contribution which can be written in the following form
\bea\label{ewt0}
E_W^{\rm vac.}=\beta\frac{\tilde{\ell}^{d-2}}{4G_N}\left(-\frac{1}{h^{d-2}}+\frac{1}{(2\ell+h)^{d-2}}\right).
\eea
The above result eq. \eqref{ewlowhigh} shows that EWCS is a monotonically decreasing function of temperature and obeys an area law even in finite temperature where the HEE shows a volume law. 

\begin{figure}
\begin{center}
\includegraphics[scale=0.72]{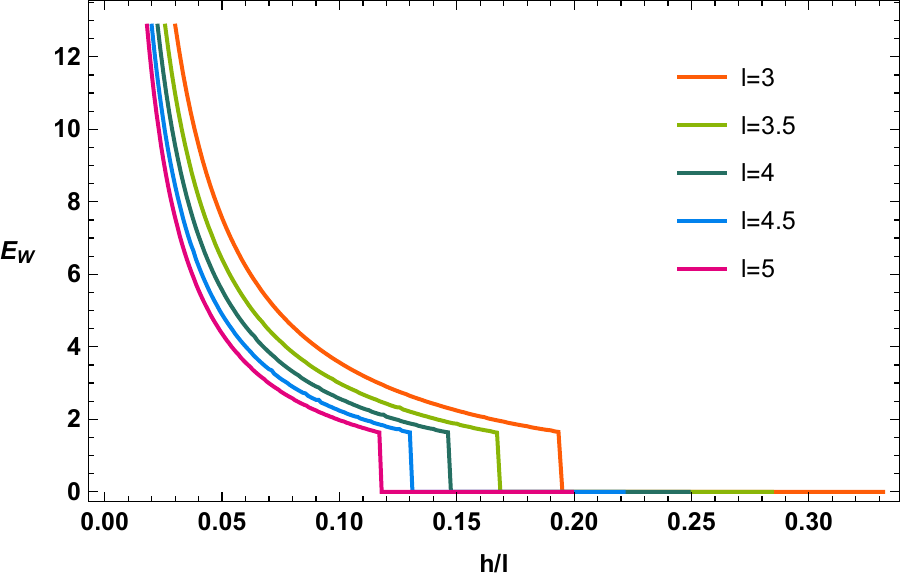}
\hspace*{0.8cm}
\includegraphics[scale=0.41]{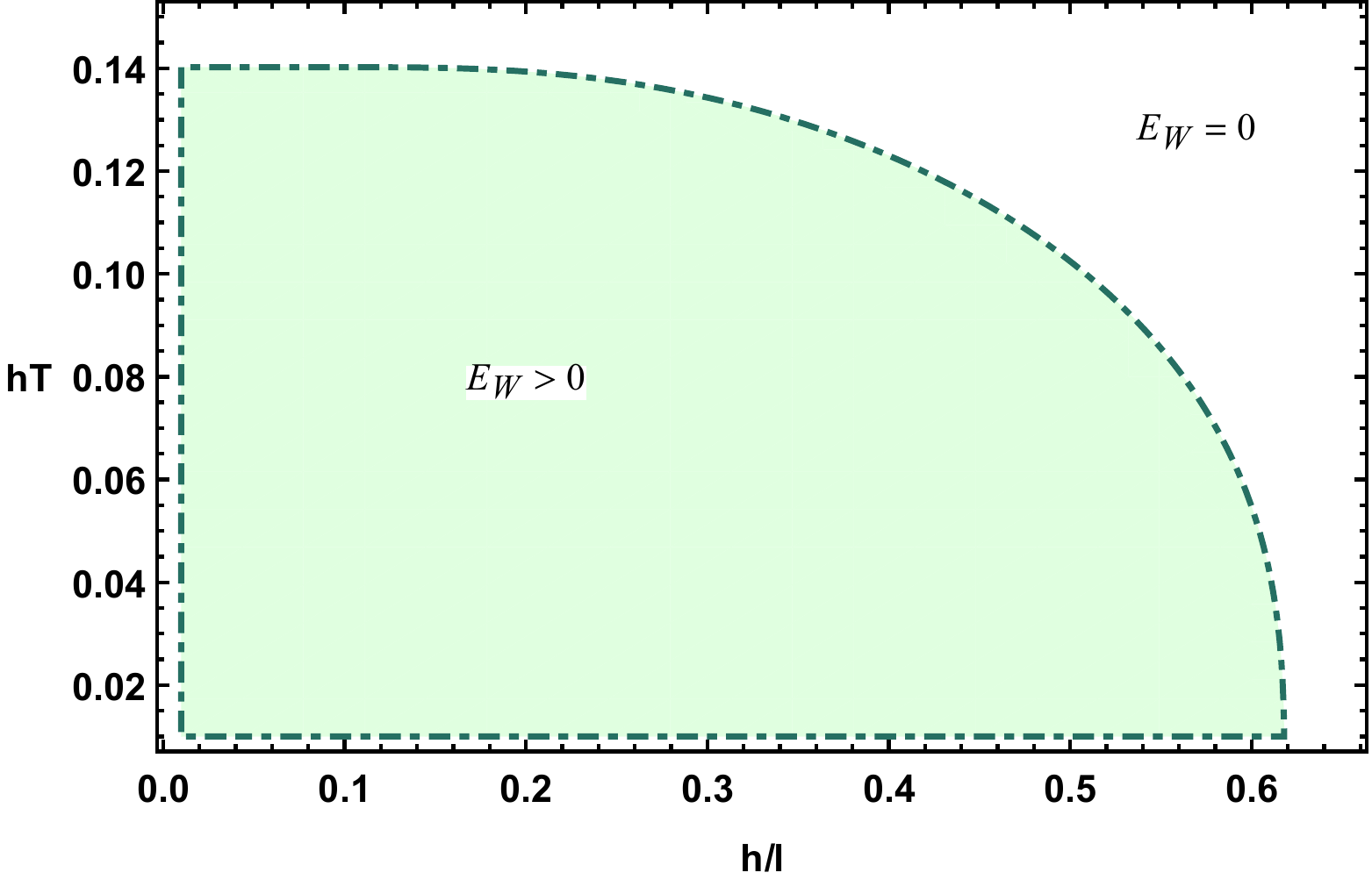}
\end{center}
\caption{\textit{Left}: $E_W$ in AdS-Schwarzschild black brane as a function of $h/\ell$ for different values of $\ell$. \textit{Right}: Parameter space for $d=3$ where $E_W$ is nonzero only in the shaded region.
 Here we have set $r_h=1$.}
\label{fig:staticEW}
\end{figure}

On the other hand, solving numerically for the turning points $r_d$ and $r_u$ using eq. \eqref{ellrt}, we can evaluate
$E_W$ in eq. \eqref{ewBB}, as shown in figure \ref{fig:staticEW} (left panel) for different values of $h$ and $\ell$. The right panel shows the two dimensional parameter space restricted by the $I(\ell,h)\geq 0$ condition which coincides with $E_W(\ell,h)\neq 0$. We can see that $E_W$ shows a discontinuous phase transition, such that $E_W=0$ when $h$ is large enough. As we mentioned before, the vanishing of EWCS results because of the disconnected configuration for the RT surfaces and the fact that in this case the corresponding entanglement wedge is empty. 

It is instructive to analyze the particular case of BTZ black holes with $d=2$, since
 $E_W$ can be determined analytically even at finite temperature. In this case the EWCS functional becomes
\bea\label{staticEW3d}
E_W=\frac{1}{4G_N}\int_{r_d}^{r_u} \frac{dr}{r\sqrt{f(r)}},\;\;\;\;\;\;\;\;f(r)=1-\frac{r^2}{r_h^2}. 
\eea
Performing the above integral, we are left with
\bea
E_W=\frac{1}{4G_N}\log\left(\frac{r_u}{r_d}\frac{1+\sqrt{f(r_d)}}{1+\sqrt{f(r_u)}}\right).
\eea
Also the relation between the width of the entangling region and the corresponding turning point at finite temperature is known \cite{Ryu:2006ef}
\bea\label{Xbb}
X=r_h\log\frac{r_h+r_t}{r_h-r_t}.
\eea
Now combining the above two equations, as well as eq. \eqref{T} for $d=2$ with zero charge, yields the following
\bea\label{ewd2bb}
E_W=\frac{c}{6}\log\frac{\tanh\frac{\pi(2\ell+h)T}{2}}{\tanh\frac{\pi hT}{2}},
\eea
where $c=\frac{3L}{2G_N}$ is the central charge. We can evaluate the zero temperature limit of the above result to find the vacuum contribution as follows
\bea\label{ewd2vac}
E_W^{\rm vac.}=\frac{c}{6}\log\frac{(2\ell+h)}{h}.
\eea

\subsection{Extremal AdS Black Brane}
\begin{figure}
\begin{center}
\includegraphics[scale=0.72]{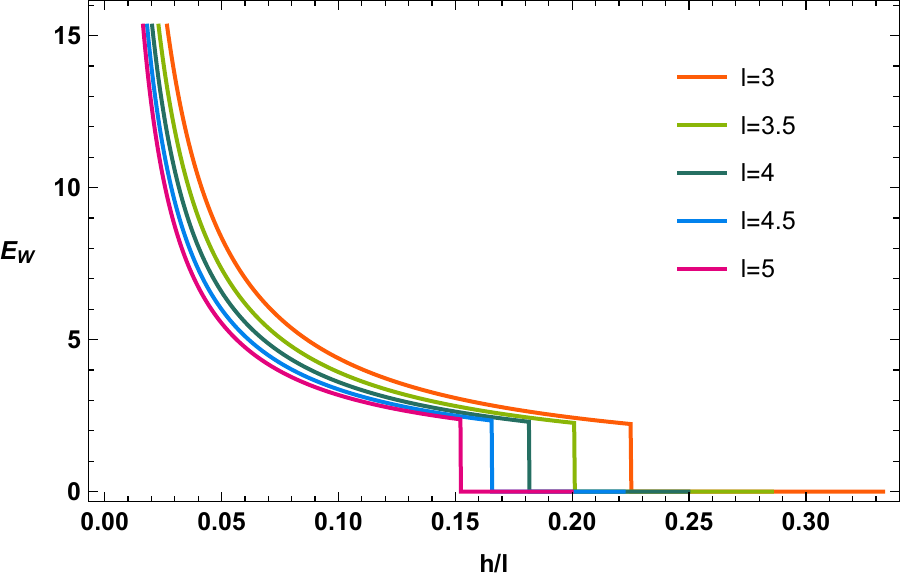}
\hspace*{0.8cm}
\includegraphics[scale=0.41]{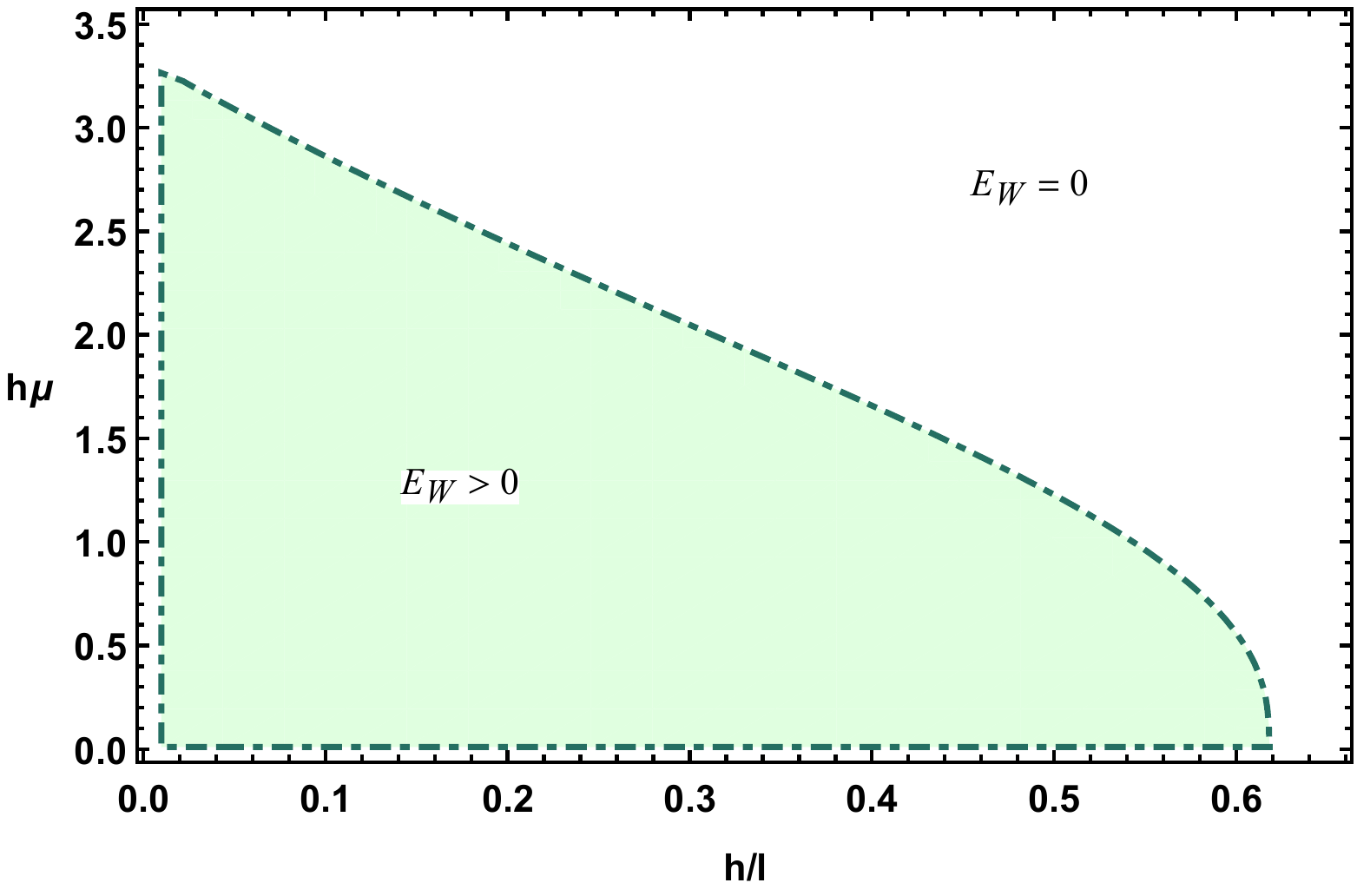}
\end{center}
\caption{\textit{Left}: $E_W$ in Extremal AdS black brane as a function of $h/\ell$ for different values of $\ell$. \textit{Right}: Parameter space for $d=3$ where $E_W$ is nonzero only in the shaded region.
 Here we have set $r_h=1$.}
\label{fig:staticEWext}
\end{figure}

In this situation, plugging eq.\eqref{fext} into eq.\eqref{staticEW}, we find 
\bea\label{extEW}
E_W=\frac{{\tilde{\ell}}^{d-2}}{4G_N}\int_{r_d}^{r_u} \frac{dr}{r^{d-1}\sqrt{1-\delta \left(\frac{r}{r_h}\right)^d+(1-\delta) \left(\frac{r}{r_h}\right)^{2d-2}}},
\eea
where we have introduced $\delta=\frac{2d-2}{d-2}$. While the above integral cannot be carried out analytically for general $d$, in a very similar manner to the analysis of the thermal correction to EWCS in \cite{BabaeiVelni:2019pkw}, we can obtain the scaling behavior of $E_W$ for extremal black branes. First, using binomial expansion we rewrite eq.\eqref{extEW} as follows 
\bea\label{extEW1}
E_W=\frac{{\tilde{\ell}}^{d-2}}{4G_N}\sum_{n=0}^{\infty}\sum_{k=0}^{n}\frac{\Gamma \left(n+\frac{1}{2}\right)\delta^{n-k}(1-\delta)^k}{\sqrt{\pi}\Gamma \left(k+1\right)\Gamma \left(n-k+1\right)}
\int_{r_d}^{r_u} \frac{dr}{r^{d-1}}\left(\frac{r}{r_h}\right)^{dn+(d-2)k},
\eea
which can be integrated to give
\bea\label{extEW2}
E_W=\frac{{\tilde{\ell}}^{d-2}}{4G_N}\sum_{n=0}^{\infty}\sum_{k=0}^{n}\frac{\Gamma \left(n+\frac{1}{2}\right)\delta^{n-k}(1-\delta)^k}{\sqrt{\pi}\Gamma \left(k+1\right)\Gamma \left(n-k+1\right)}
\frac{r^{dn+(d-2)(k-1)}}{(dn+(d-2)(k-1))r_h^{dn+(d-2)k}}\Big|_{r_d}^{r_u}.
\eea
Further, the relation between the position of the turning point $r_t$ and the strip width $X$ can be written as follows \cite{Kundu:2016dyk}
\bea\label{ellrtext}
X=r_t\sum_{n=0}^{\infty}\sum_{k=0}^{n}\frac{\Gamma \left(n+\frac{1}{2}\right)\delta^{n-k}(1-\delta)^k}{\Gamma \left(k+1\right)\Gamma \left(n-k+1\right)}\frac{\Gamma \left(\frac{dn+(d-2)k+d}{2(d-1)}\right)}{\Gamma \left(\frac{dn+(d-2)k+2d-1}{2(d-1)}\right)}\left(\frac{r_t}{r_h}\right)^{dn+(d-2)k}.
\eea
Now we would like to invert this formula to write eq. \eqref{extEW2} in terms of the boundary quantities, $h$, $2\ell+h$ and $\mu$. A similar
derivation to the one presented for AdS-Schwarzschild black brane holds in this case. Again, to perform an exact estimation, we will focus on the behavior of EWCS in small and large chemical potential limits. As demonstrated in\cite{Kundu:2016dyk}, considering small chemical potential with respect to the separation scale corresponds to $h\ll \mu^{-1}\ll \ell$. Once again, one might also regard the $h\ll \ell\ll \mu^{-1}$ case where we have small chemical potential with respect to both the subregion sizes and the separation between them. Next, we note that $h\ll \mu^{-1}\ll \ell$ limit corresponds to $r_d\ll r_h$ and $r_u\rightarrow r_h$, while for $h\ll \ell\ll \mu^{-1}$ we have $r_d, r_u\ll r_h$. Further, using eqs. \eqref{extEW2} and \eqref{ellrtext} one can find that the small and large chemical potential expansion of EWCS for $d>3$ is given by
\bea\label{ewchem}
E_W\sim \Bigg\{ \begin{array}{rcl}
&E_W^{\rm vac.}+\alpha' \frac{\tilde{\ell}^{d-2}}{G_N}\ell(\ell+h)\mu^{d}+\cdots&\,\,\,h\ll \ell \ll \mu^{-1}\\
&\frac{\tilde{\ell}^{d-2}\mu^{d-2}}{4G_N}\left(-\frac{\beta'}{(h\mu)^{d-2}}+\left(\frac{4\pi}{d}\right)^{d-1}\gamma'+\lambda'(h\mu)^2\right),
 &\,\,\,h\ll \mu^{-1}\ll \ell 
\end{array}.
\eea
where $\alpha'$, $\beta'$, $\gamma'$ and $\lambda'$ are some constants that depends on $d$. This result shows that EWCS is a monotonically increasing function of $\mu$ and obeys an area law even in finite chemical potential. Again, solving numerically for the turning points using eq. \eqref{ellrtext}, we can find $E_W$, as shown in figure \ref{fig:staticEWext} (left panel) for different values of $h$ and $\ell$. Also the right panel presents the two dimensional parameter space restricted by the $E_W(\ell,h)\neq 0$ condition. We note again that $E_W$ shows a discontinuous phase transition, such that $E_W=0$ when $h$ is large enough.

\section{EWCS in Vaidya backgrounds: $2+1$ dimensions }\label{3dim}
In this section, we study the time evolution of EWCS by considering the case where $d=2$ and the final equilibrium state is given by the BTZ black hole. First, we provide a numerical analysis and examine the various regimes in the growth of EWCS in the thin shell limit. 
Next, we will show that $\Sigma$ is a geodesic whose length can be expressed analytically in closed form, which enables us to directly extract its scaling behavior in various regimes. 
\subsection{Numerical analysis}
\begin{figure}
\begin{center}
\includegraphics[scale=0.34]{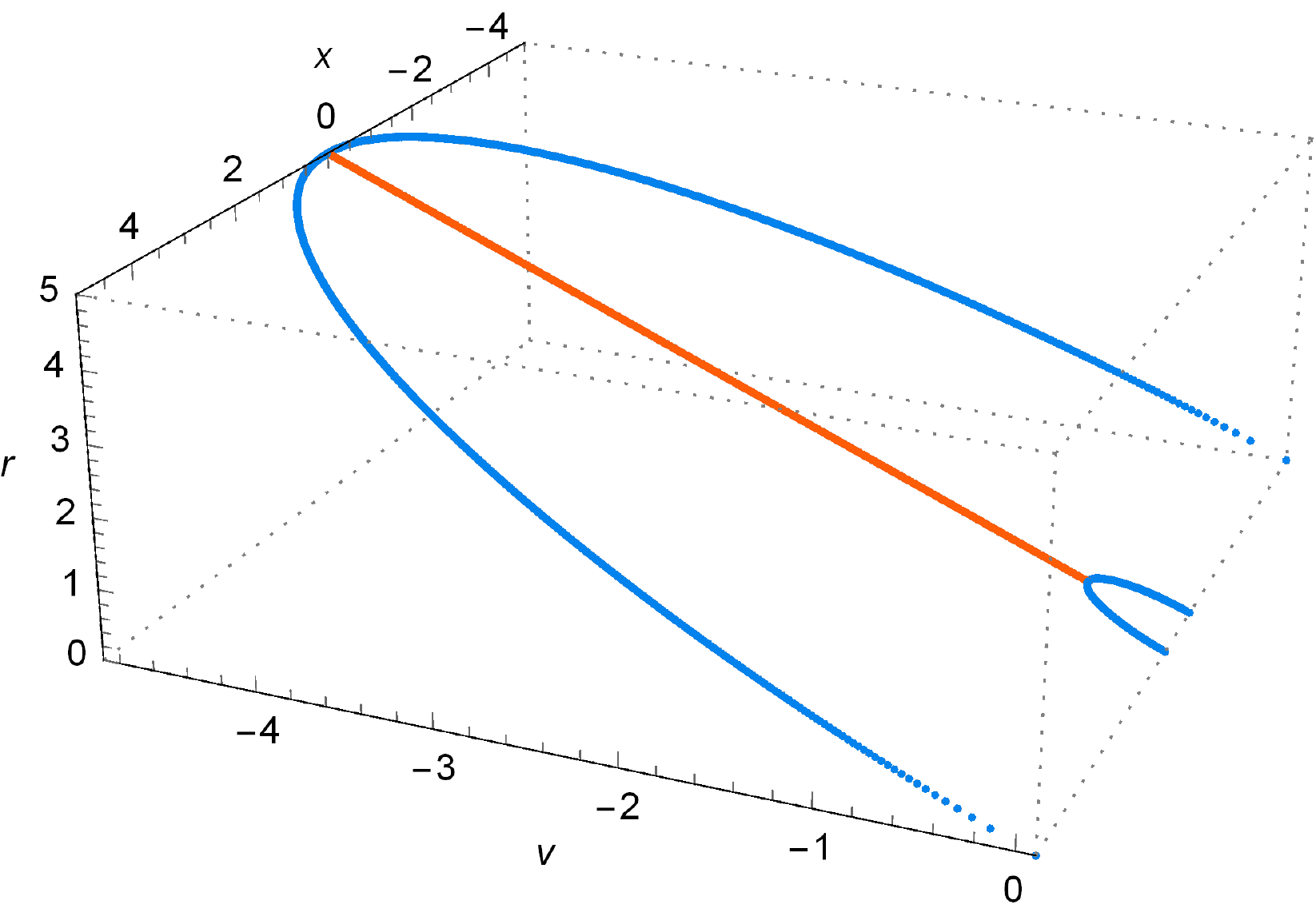}
\hspace*{0.6cm}
\includegraphics[scale=0.4]{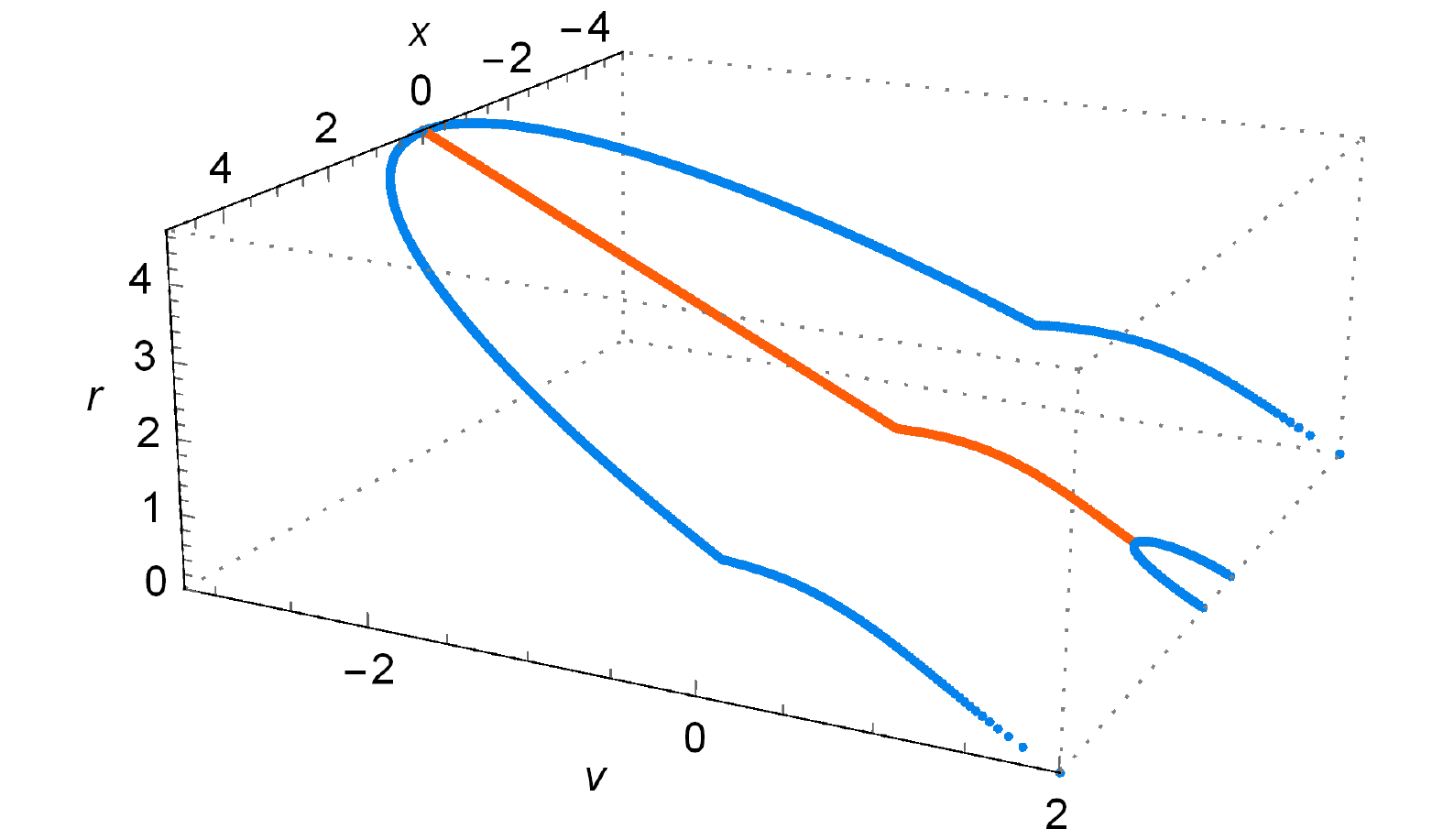}
\hspace*{0.6cm}
\includegraphics[scale=0.37]{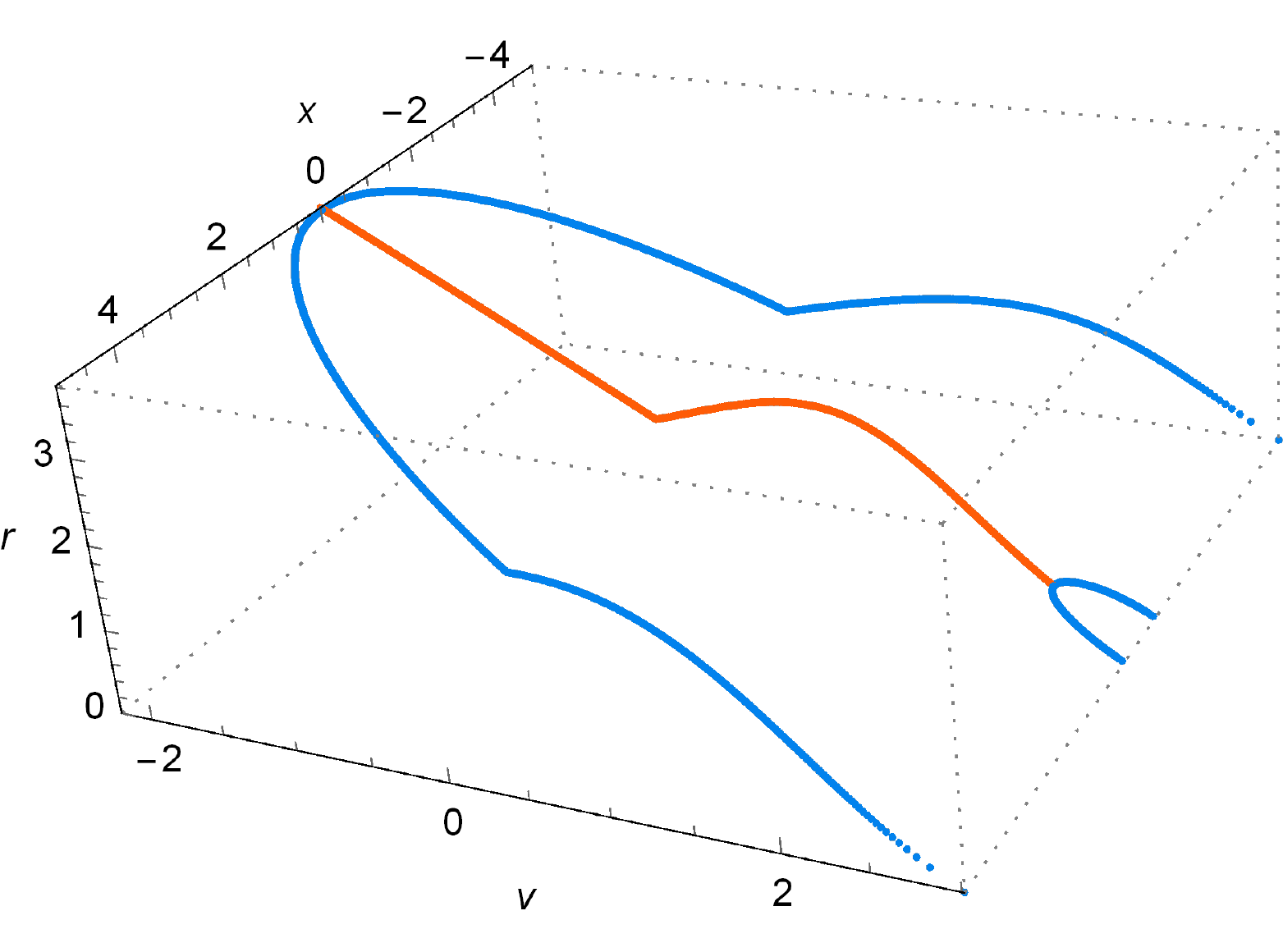}
\hspace*{0.6cm}
\includegraphics[scale=0.4]{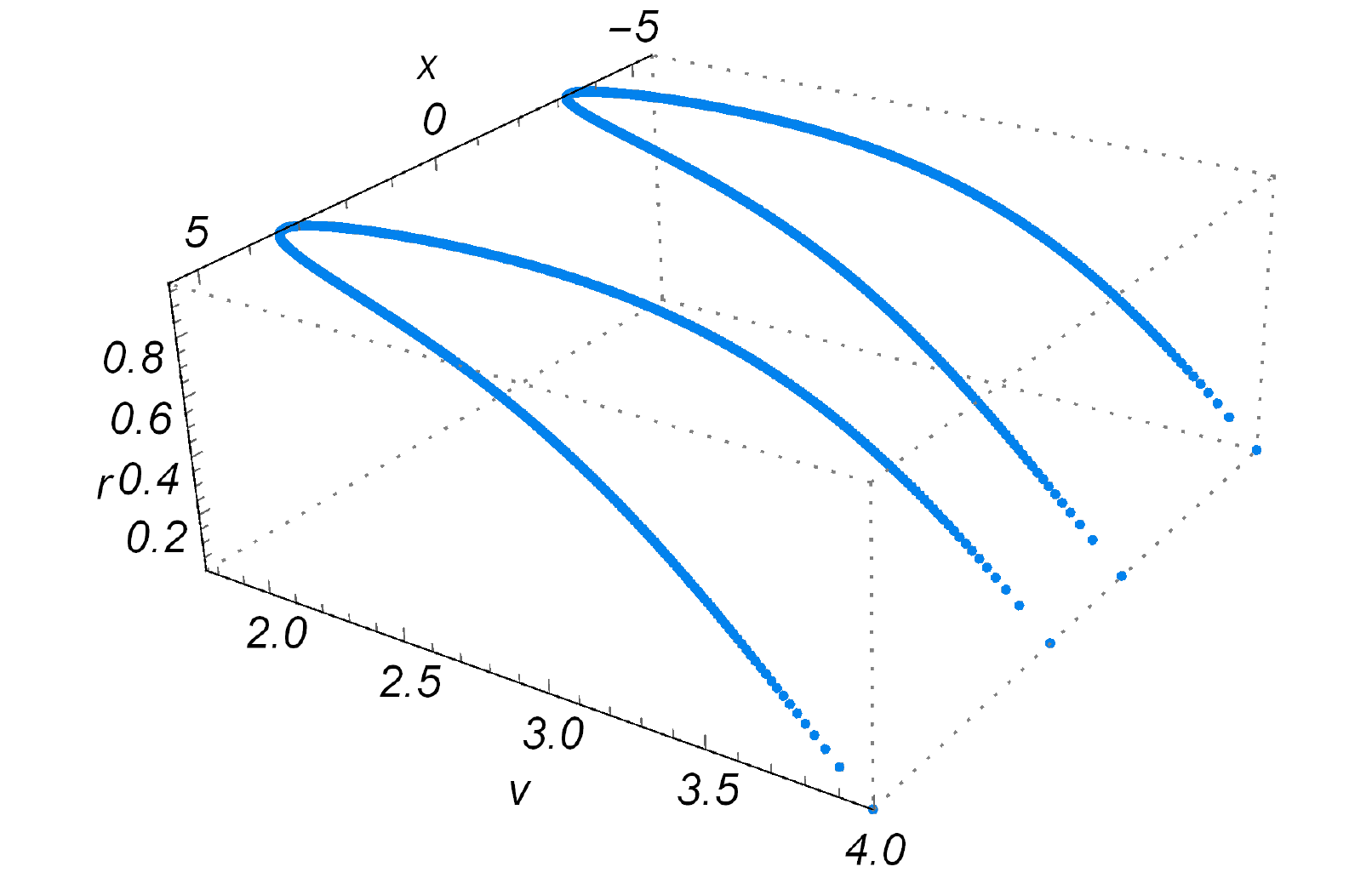}
\end{center}
\caption{Configuration of the minimal hypersurfaces for computing HEE (cyan) and EWCS (orange) at different boundary times: $t=0.1, 2, 3, 4$. Here we consider $h=1$ and $\ell=4.5$. In this case the disconnected configuration is favored at late times when $\Sigma$ becomes empty and $E_W$ saturates to zero.}
\label{fig:HRTEW}
\end{figure}
We start by evaluating $E_W(t)$, defined in eq. \eqref{funcEWCS}, numerically for several values of $h, \ell$ and $T$ . We will consider subsystems consisting of equal width intervals as depicted in figure \ref{fig:regions}. For simplicity, we set $r_h=1$ and work with the rescaled quantity $\tilde{E}_W=4G_N E_W$ throughout the following. 
We will focus on thermal quench where the corresponding geometry is given by eq. \eqref{vaidyametric} with $m(v)$ is given by eq. \eqref{mq} and $q(v)=0$. As we mentioned before we consider the thin shell regime where $v_0\rightarrow 0$ and the corresponding mass function behaves like a step function. To do so, we consider $v_0=10^{-3}$ throughout the following. Note that the EWCS is nontrivial only for connected configurations, i.e., $\Gamma_{A\cup B}=\Gamma_{2\ell+h}+\Gamma_{h}$, and vanishes for disconnected ones, i.e., $\Gamma_{A\cup B}=2\Gamma_{\ell}$ ,  when $\Sigma$ becomes empty. The extremal hypersurfaces can be found by solving eqs. \eqref{EOMHEE} and \eqref{EOMEWCS}. We show the full profile of these hypersurfaces for a specific value of $h$ and $\ell$ at different values of boundary times in figure \ref{fig:HRTEW}. In this case, the disconnected configuration is favored at late times and $E_W$ saturates to zero. 
Note that the most straightforward way to choose the minimal area configuration is by comparing the corresponding entanglement entropies. Another way is to compute the mutual information noting that in the disconnected phase the HMI vanishes. 
\begin{figure}
\begin{center}
\includegraphics[scale=1.05]{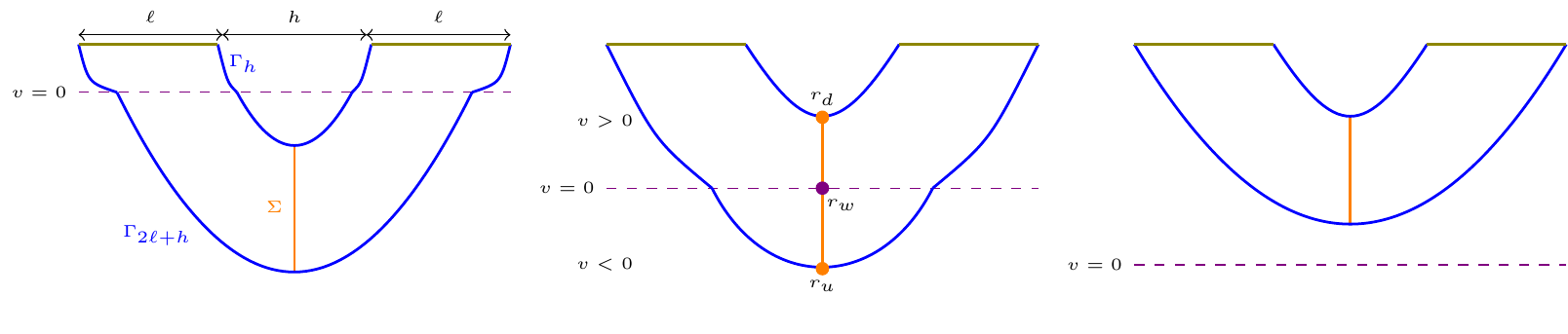}
\end{center}
\caption{Schematic configurations for HRT ($\Gamma$) and EWCS ($\Sigma$) geodesics corresponding to symmetric entangling regions on the boundary. Outside the null shell (indicated in dashed violet), i.e, $v>0$, the geodesics propagate in an AdS black brane geometry, while inside, i.e., $v<0$, they propagate in a pure AdS geometry. The null shell refracts the geodesics. \textit{Left}: While at early times, the shell lies in the UV part of the geometry and $\Gamma$'s cross the shell near the boundary, $\Sigma$ does not intersect the shell. \textit{Middle}: During intermediate stages of time evolution $\Gamma_{2\ell+h}$ and $\Sigma$ cross the null shell. \textit{Right}: At late time $\Gamma$'s and $\Sigma$ do not intersect with the shell and lie entirely in black brane geometry.}
\label{fig:regions1}
\end{figure}

Regarding the evolution of EWCS and assuming that  the connected configuration is always favored for any boundary time, there are three different scaling regimes\footnote{Note that for the case where the disconnected configuration is favored at late times, only the late time behavior should be suitably modified and the last panel in Fig. \ref{fig:regions1} must be replaced with a disconncted configuration.} (see Fig. \ref{fig:regions1}) (i) at early time $\Sigma$ does not reach the shell and lies entirely
in AdS geometry, (ii) during intermediate stage of time evolution, $\Sigma$ crosses the shell at $r_w$ such that $r_d<r_w<r_u$, and (iii) at late time $\Sigma$ lies entirely in black brane geometry. 
\begin{figure}[h]
\begin{center}
\includegraphics[scale=0.415]{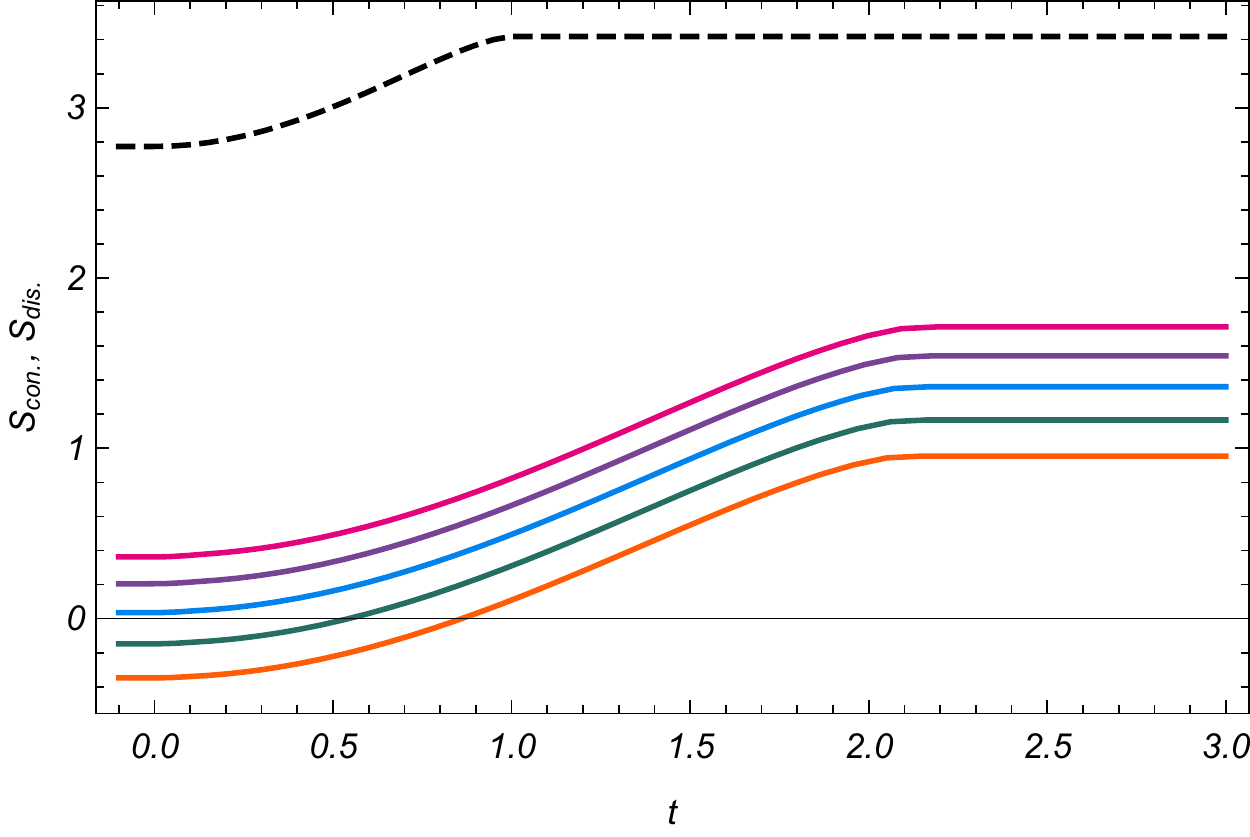}
\hspace*{0.05cm}
\includegraphics[scale=0.59]{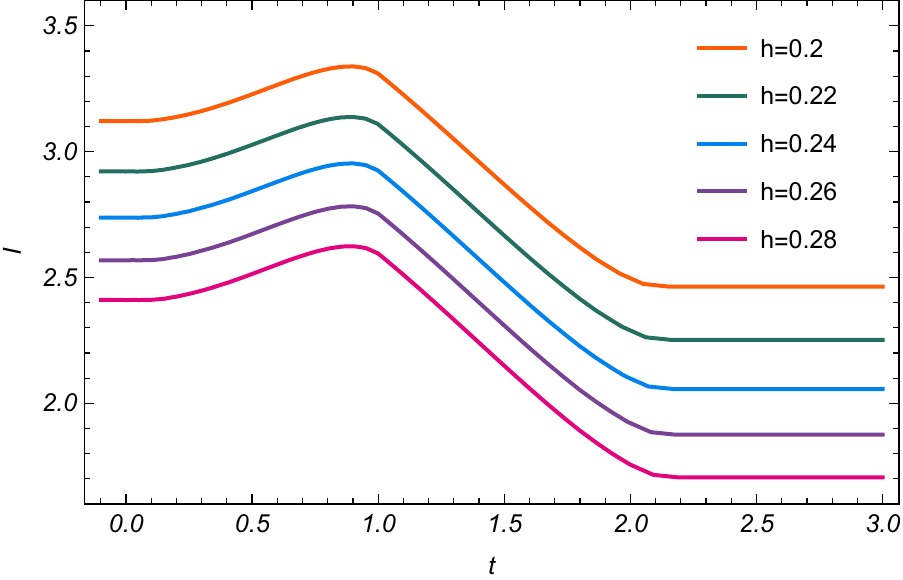}
\hspace*{0.05cm}
\includegraphics[scale=0.415]{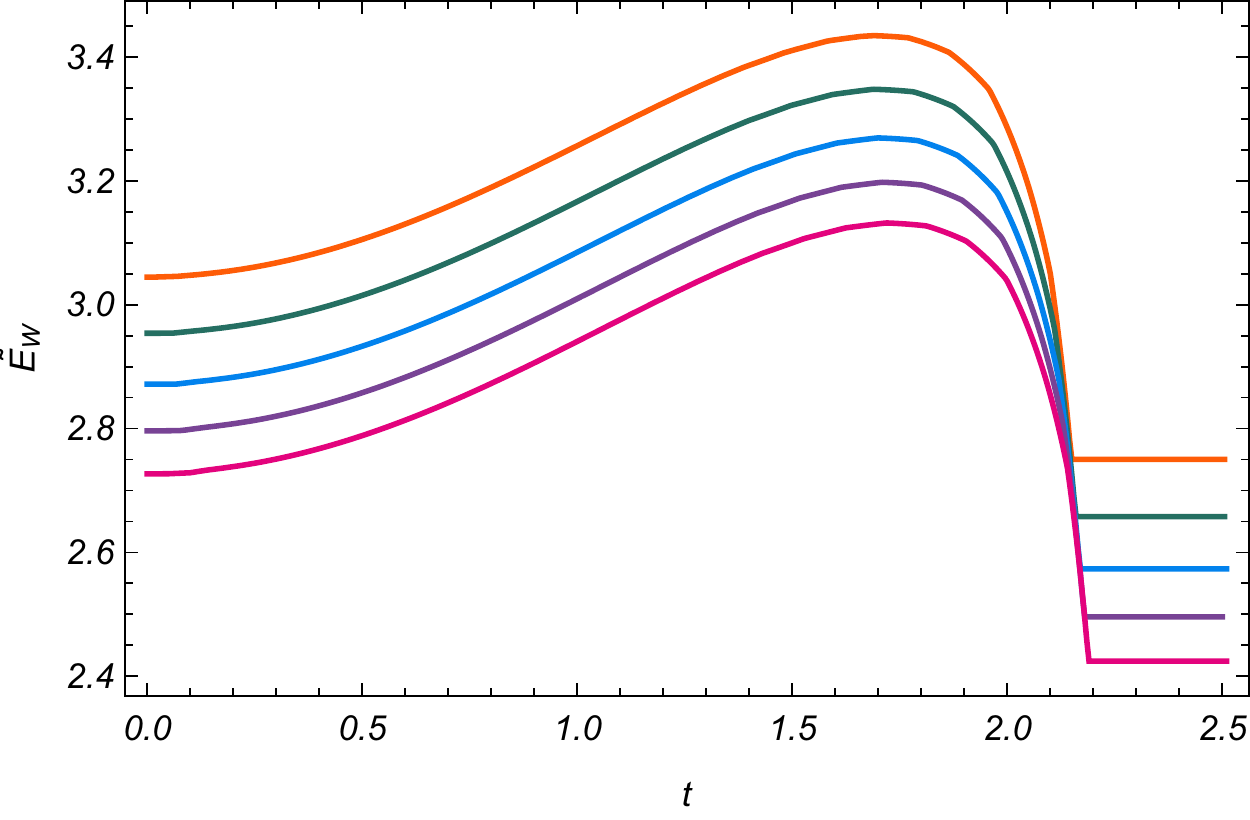}
\end{center}
\caption{\textit{Left}: The HEE for connected (solid) and disconnected (dashed) configurations for different values of the separation between subregions. The connected configuration is always favored for any boundary time. \textit{Middle}: The HMI as a function of time which is positive during the entire evolution. \textit{Right}: The EWCS as a function of boundary time which saturates to a finite value. Here we have set $\ell=2$.}
\label{fig:2d-1}
\end{figure}

\begin{figure}[h]
\begin{center}
\includegraphics[scale=0.416]{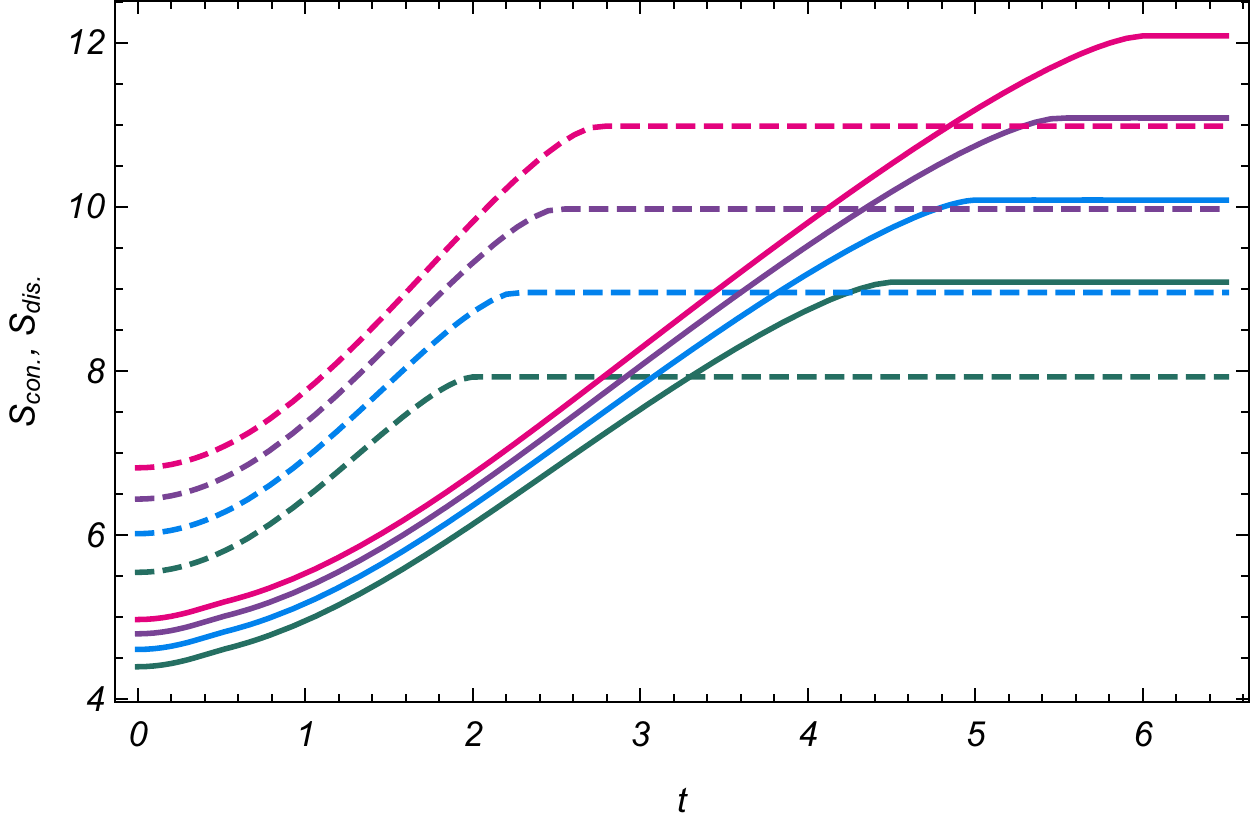}
\hspace*{0.05cm}
\includegraphics[scale=0.416]{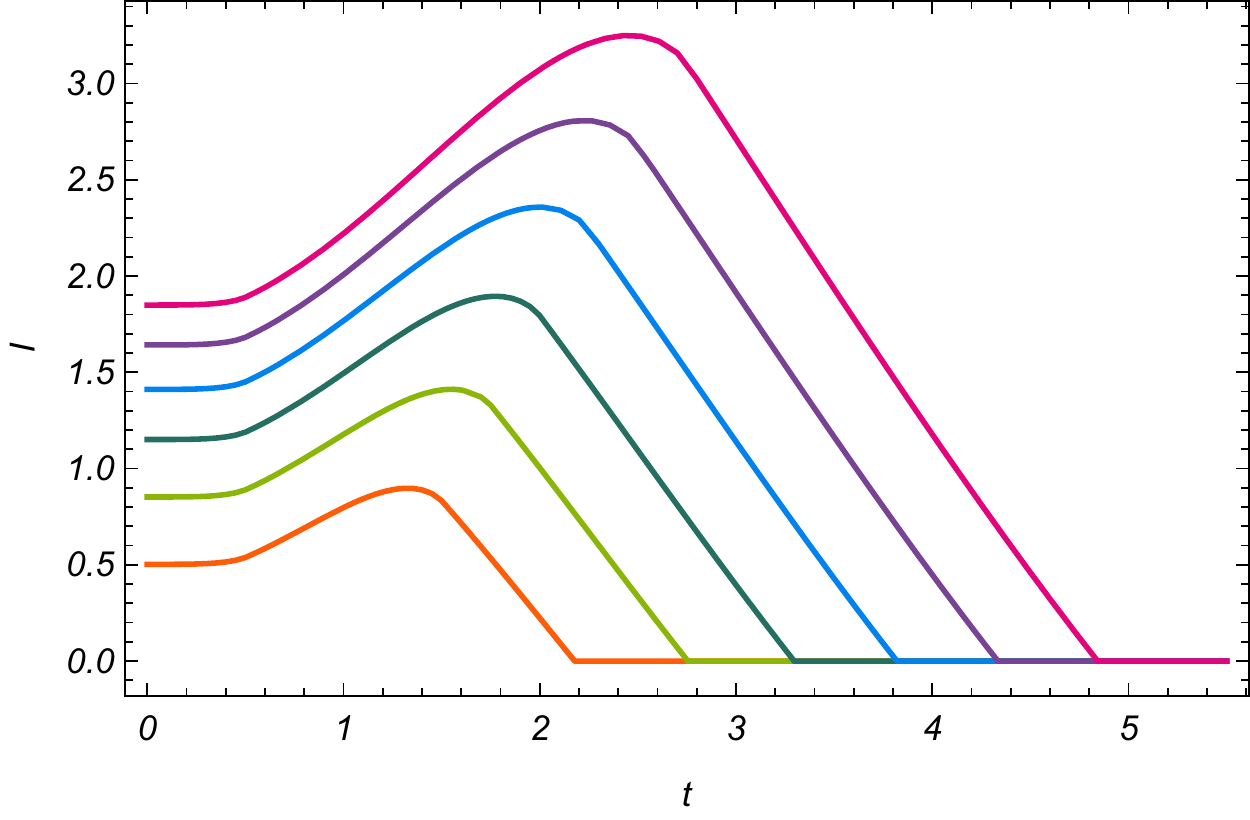}
\hspace*{0.05cm}
\includegraphics[scale=0.58]{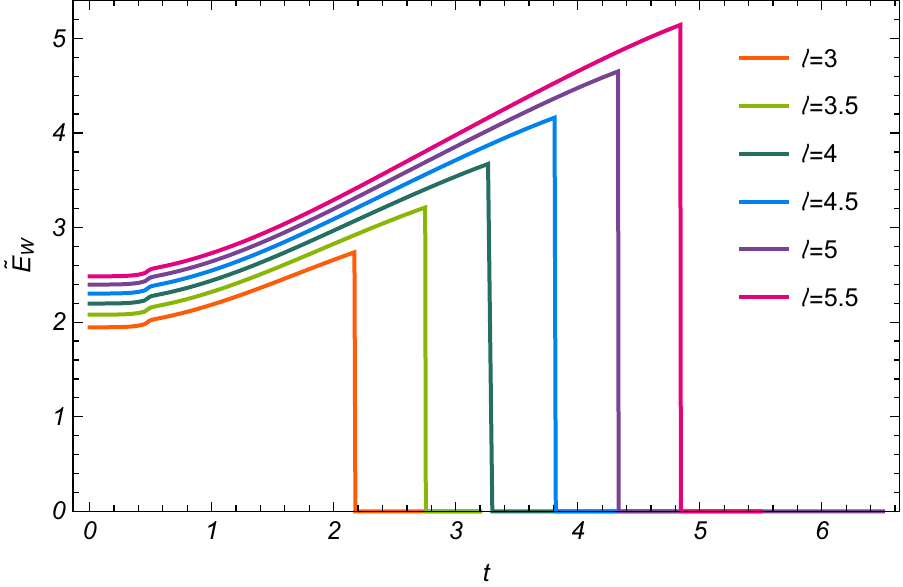}
\end{center}
\caption{\textit{Left}: The HEE for connected (solid) and disconnected (dashed) configurations for different values of $h$. At late times, the disconnected configuration is favored. \textit{Middle}: The HMI as a function of time which vanishes at late times. \textit{Right}: The EWCS as a function of boundary time which saturates to zero. Here we have set $h=1$.}
\label{fig:2d-2}
\end{figure}

In figures \ref{fig:2d-1} and \ref{fig:2d-2} we show the competition between connected and disconnected configurations, the corresponding HMI and EWCS for specific values of $h$ and $\ell$. 
In figure \ref{fig:2d-1}, we show various boundary quantities for the case where both the subregions width and separation between them are small, i.e., $h< \ell < T^{-1}$. In this example, we fix $\ell$ and consider different values for $h$. In the left panel, we compare the values of $S_{\rm con.}(t)$ and $S_{\rm dis.}(t)$ to see which configuration is minimal during the evolution. Our numerical results make it clear that in this case the connected configuration is always favored for any time. The middle panel demonstrates the evolution of HMI which is always nonzero, irrespective of the boundary time. 
In the right panel we show $E_W(t)$ for the same values of the parameters. At early times, i.e., $t\ll h$, the EWCS starts at the same value of the AdS geometry given by eq. \eqref{ewd2vac}, then at $t\sim \mathcal{O}(h)$ quickly deviates from the vacuum value and approaches a regime of linear growth. Based on these plots, we observe a period of time during which
the growth of the $E_W$ is quadratic. We will examine this observation further in the following. Note that in the period of linear growth, the slope seems more or less the same independent of $h$. This regime in fact persists all the way up to $t\sim \mathcal{O}(\ell+h)$ where $E_W$ reaches its maximum value. Further, at late times, $E_W$ decreases and very quickly saturates to a constant value corresponding to the BTZ geometry given by eq. \eqref{ewd2bb}.

We present the time dependence of the EWCS for the case of $h < T^{-1} < \ell$ in figure \ref{fig:2d-2}. 
The left panel shows the competition between the contribution to HEE due to the connected and disconnected configurations. Based on this figure, although the connected configuration has the minimal area at early time, the late time behavior is governed by the disconnected configuration. The critical time when this transition happens is approximately given by $t\sim \mathcal{O}(\ell-h)$. We show $E_W(t)$ for the same values of the parameters in the right panel. Once again, at early times the EWCS starts growing quadratically from the vacuum value and approaches the linear growth regime. It seems that, in this regime, the slope is independent of $\ell$. Finally, at late times, $E_W$ displays a discontinuous transition and immediately saturates to zero where the saturation time is approximately $t_s\sim \mathcal{O}(\ell-h)$. 

\begin{figure}
\begin{center}
\includegraphics[scale=0.82]{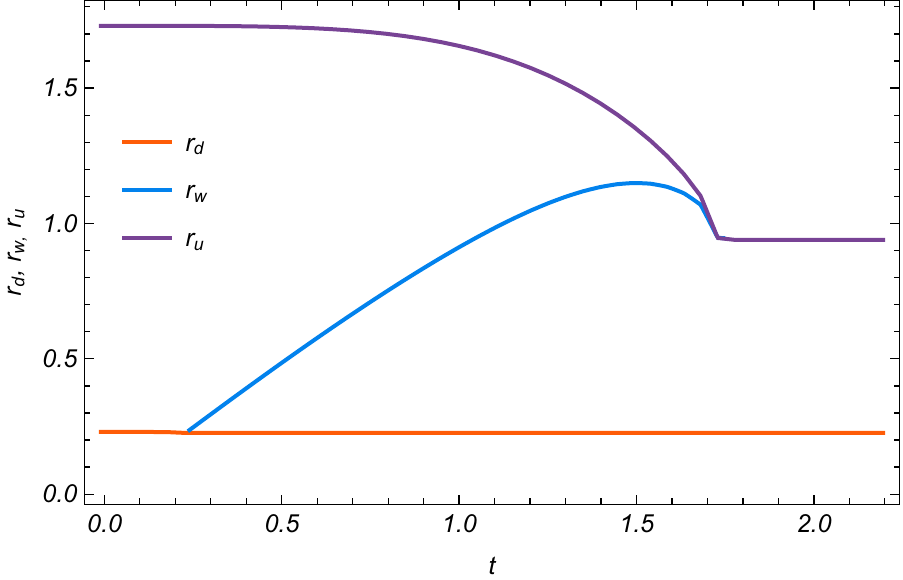}
\hspace*{1cm}
\includegraphics[scale=0.8]{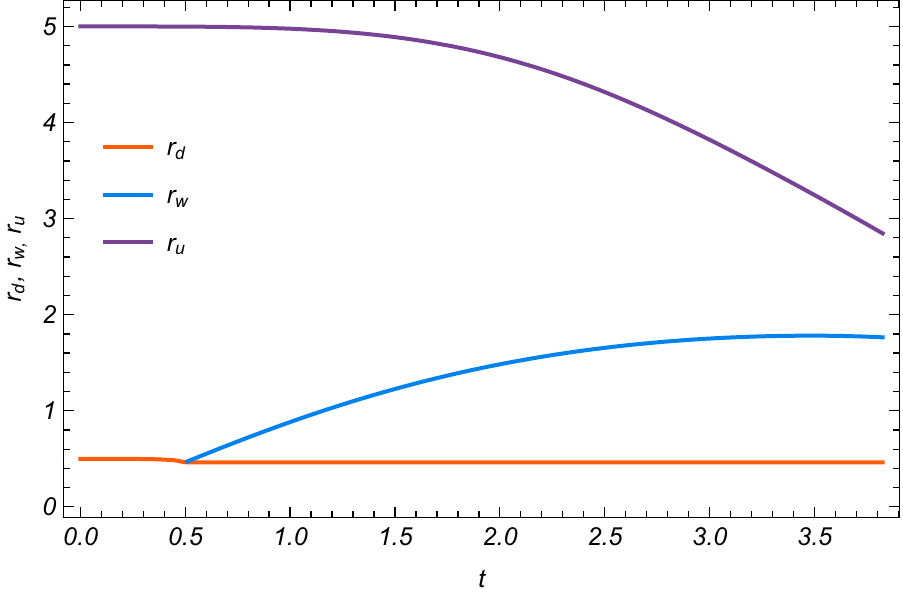}
\end{center}
\caption{Time evolution of $r_d, r_w$ and $r_u$. Clearly, $r_w$ always obeys the bound $r_d\leq r_w \leq r_u$. \textit{Left}: For $h=0.46$ and $\ell=1.5$ the connected configuration is always favored for any boundary time. \textit{Right}: For $h=1$ and $\ell=4.5$ the disconnected configuration is favored at late times. Here the transition between connected and disconnected configurations happens at $t\sim 3.82$. }
\label{fig:rwrdrunum}
\end{figure}

\begin{figure}
\begin{center}
\includegraphics[scale=0.6]{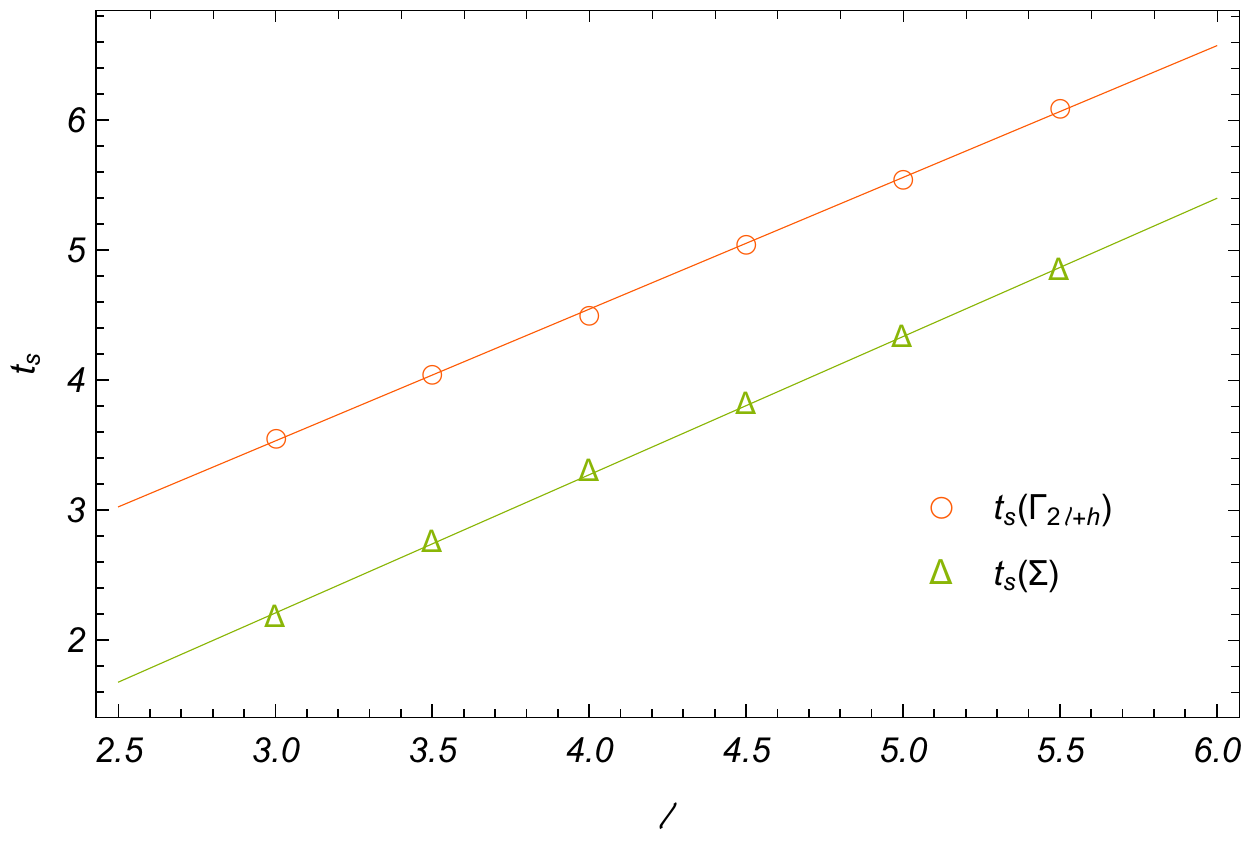}
\end{center}
\caption{Comparison of the saturation time of the EWCS and the saturation time of the HEE corresponding to $\Gamma_{2\ell+h}$ when the disconnected configuration is favored at late times for different values of $\ell$. Clearly, $t_s(\Gamma_{2\ell+h})$ is always greater than $t_s(\Sigma)$. Here we have set $h=1$. }
\label{fig:saturationtime}
\end{figure}

To conclude this section let us comment on the essential role of the minimal hypersurface $\Gamma_{2\ell+h}$. As noted above, one can identify the three positions, i.e., $r_d, r_u$ and $r_w$ which are important in studying the evolution of EWCS. We can also consider the time dependence of these points, as shown in figure \ref{fig:rwrdrunum} for specific values of $h$ and $\ell$. In the left panel, we show the results for $h=0.46$ and $\ell=1.5$ where the connected configuration is always favored for any boundary time and the resultant $E_W$ continuously saturates to the final equilibrium value given by eq. \eqref{ewd2bb}. The right panel shows the results for $h=1$ and $\ell=4.5$. In this case the disconnected configuration is favored at late times. To see this, note that $I(\ell, h)>0$ condition in $d=2$ yields $h<h_{\rm crit.}$, where the critical separation is given by
\bea\label{hcritBTZ}
h_{\rm crit.}=\cosh^{-1}\left(2\cosh(\ell)-1\right)-\ell.
\eea
Also note that for both cases $r_w$ always obeys the bound $r_d\leq r_w \leq r_u$ as expected.
These plots show that the main behavior in the evolution of $\Sigma$ depends on $r_u$ which is the corresponding turning point of $\Gamma_{2\ell+h}$ while $r_d$ and the shape of $\Gamma_{h}$ are fixed and do not influence the time dependence of $E_W(t)$ for the present symmetric case. 
However, we should also remark that in general, the saturation time of the EWCS does not follow the saturation time of the HEE corresponding to $\Gamma_{2\ell+h}$. Although when the connected configuration is favored at late times these time scales coincide, one should take into account that for some boundary subregions the minimal configuration is given by the disconnected solution at late times. In the latter case the saturation time of EWCS is smaller than the saturation time of the HEE associated with $\Gamma_{2\ell+h}$. We compare these two time scales as a function of $\ell$ in figure \ref{fig:saturationtime}.

\subsection{Analytic treatment}

When the final equilibrium state is given by the BTZ black brane, most of the expressions can be evaluated analytically. In particular, in the thin shell approximation, we demonstrate that the problem admits semi-analytic solution. In fact, this provides
a check on our numerical results and also allow us to derive in detail several general features in the time evolution of EWCS.

As we explain in the previous section, the evolution of EWCS can be divided into three different scaling regimes (see Fig. \ref{fig:regions1}). In cases (i) and (iii) corresponding to the early and late time static geometries, the EWCS lies entirely on a constant time slice and we can use the previous expressions derived in sec. \ref{sec:adsbb} to find $E_W$. On the other hand, in case (ii) the part of $\Sigma$ that is inside the shell is given by the geodesic in the pure AdS geometry, and the part of it that is outside the shell is given by the geodesic in the pure black brane geometry. In this case, the geodesic gets refracted at the null shell and it does not have to be in a constant time slice. In the following we focus on case (ii) which is more involved.
Using eq. \eqref{vaidyametric}, we write the metric in zero charge limit as
\bea
ds^2=\frac{1}{r^2}\left[-f(r,v)dv^2-2dv dr +dx^2\right],\;\;\;\;f(r,v)=1-m(v) r^2.
\eea
Further, we consider a thin null shell such that $m(v)=\frac{\theta(v)}{r_h^{2}}$ and
\bea
f(r,v)=\Bigg\{ \begin{array}{rcl}
&1\equiv f_{a}&\,\,\,v<0\;\;\;\\
& 1-\frac{r^2}{r_h^2}\equiv f_{b}&\,\,\,v>0
\end{array}.
\eea
Note we have used the subscript $a$ ($b$) to refer to quantities on the AdS (black brane) side of the null shell. In this case, using eq. \eqref{vt}, the boundary time reads
\bea\label{vtd3}
t=v_b-\frac{r_h}{2}\log\frac{r_h-r}{r_h+r}.
\eea
The geodesic equations in the black brane geometry become\cite{Balasubramanian:2011ur}
\bea\label{geodesics}
\frac{f_b(r)\dot{t}}{r^2}=\frac{\mathcal{Q}_b}{r_h},\nonumber\\
1=-\frac{f_b(r)}{r^2}\dot{t}^2+\frac{\dot{r}^2}{r^2f_b(r)},
\eea
where $\mathcal{Q}_b$ is some integration constant, $\dot{} \equiv \frac{d}{d\lambda}$ and $\lambda$ parametrizing the geodesic length.  Note that we consider geodesics that lie on $x=0$ slice, so comparing to \cite{Balasubramanian:2011ur} there is only one integration constant. Combining these two equations together yields
\bea
\dot{r}^2=r^2\left(1-\frac{r^2}{r_h^2}+\frac{r^2\mathcal{Q}_b^2}{r_h^2}\right),
\eea
which can be solved as follows
\bea
r(\lambda)^2=\frac{4r_h^2}{\left(e^{\lambda}+\left(1-\mathcal{Q}_b^2\right)e^{-\lambda}\right)^2}.
\eea
Now using the above result we can solve eq. \eqref{geodesics} for $t(r)$ to find
\bea
t_{\pm}(r)=c_{\pm}+\frac{r_h}{2}\log\left|\frac{r_h^2-(\mathcal{Q}_b+1)r^2\pm r_h\sqrt{r_h^2+(\mathcal{Q}_b^2-1)r^2}}{r_h^2+(\mathcal{Q}_b-1)r^2\pm r_h\sqrt{r_h^2+(\mathcal{Q}_b^2-1)r^2}}\right|,
\eea
where $c_{\pm}$ are integration constants. Hereafter, we only consider the $+$ branch of the geodesics without loss of generality. Next, using eq. \eqref{vtd3} for $v_b(r)$, we obtain
\bea\label{BBvr}
v_{b}(r)=c_++\frac{r_h}{2}\log\left(\frac{r_h-r}{r_h+r}\frac{r_h^2-(\mathcal{Q}_b+1)r^2+ r_h\sqrt{r_h^2+(\mathcal{Q}_b^2-1)r^2}}{r_h^2+(\mathcal{Q}_b-1)r^2+r_h\sqrt{r_h^2+(\mathcal{Q}_b^2-1)r^2}}\right),
\eea
which can be rewrite as follows
\bea\label{BBvr1}
v_{b}(r)=c_{b}+r_h\log\frac{\mathcal{Q}_b r_h-\sqrt{r_h^2+(\mathcal{Q}_b^2-1)r^2}}{\mathcal{Q}_b(r+r_h)}.
\eea
Note that the expression for the part of geodesic that lies in the pure AdS geometry inside the null shell can be obtained from the $r_h\rightarrow \infty$ limit of eq. \eqref{BBvr} as follows
\bea\label{adsvr}
v_a(r)=c_a-r-\frac{\sqrt{1+\mathcal{Q}_a^2r^2}}{\mathcal{Q}_a},
\eea
where in doing this, we should scale $\mathcal{Q}_b$ with $r_h$ at the same time because in \eqref{geodesics} we have defined the integration constant with an extra factor of horizon radius. The integration constants , i.e., $c_a$ and $c_b$, will be fixed by setting $v_a(r_u)=v_u$ and $v_b(r_d)=v_d$, respectively. Imposing these conditions, we have
\bea\label{cacb}
c_a=r_u+v_u+\frac{\sqrt{1+\mathcal{Q}_a^2r_u^2}}{\mathcal{Q}_a},\;\;\;\;\;\;c_b=v_d-r_h\log\frac{\mathcal{Q}_b r_h-\sqrt{r_h^2+(\mathcal{Q}_b^2-1)r_d^2}}{\mathcal{Q}_b(r_d+r_h)}.
\eea
On the other hand, $\mathcal{Q}_a$ and $\mathcal{Q}_b$ can be found using the matching conditions at the null shell. Denoting value of $r$ at the intersection of $\Sigma$ and the null shell $v=0$ as $r_w$, we  note that $v'(r)$ will be discontinuous at this point because of the refraction condition noted above. To find the matching condition for the derivative we integrate the equation of motion across the null shell which reads
\bea\label{match2}
\frac{dr}{dv}|_{a}-\frac{dr}{dv}|_{b}=-\frac{1}{2}\frac{r_w^2}{r_h^2}.
\eea
Solving the above condition, we obtain
\bea\label{Qb}
\mathcal{Q}_b=\pm\frac{(r_w^2-2r_h^2)\mathcal{Q}_a+r_w\sqrt{1+\mathcal{Q}_a^2r_w^2}}{2r_h}.
\eea
On the other hand, at the intersection of $\Sigma$ and the null shell we have
\bea\label{match1}
v_a(r_w)=0=v_b(r_w),
\eea
which means that $v$ remains continuous along $r=r_w$. The above conditions can be solved analytically in closed form as follows
\bea\label{Qa}
\mathcal{Q}_a^2=\frac{4(v_u+r_u-r_w)^2}{v_u(v_u-2r_w)(v_u+2r_u)(v_u+2r_u-2r_w)},
\eea
\bea\label{rweq}
\frac{\mathcal{Q}_b r_h-\sqrt{r_h^2+(\mathcal{Q}_b^2-1)r_d^2}}{\mathcal{Q}_b r_h-\sqrt{r_h^2+(\mathcal{Q}_b^2-1)r_w^2}}\frac{r_h+r_w}{r_h+r_d}=e^{\frac{v_d}{r_h}}.
\eea

\begin{figure}
\begin{center}
\includegraphics[scale=0.845]{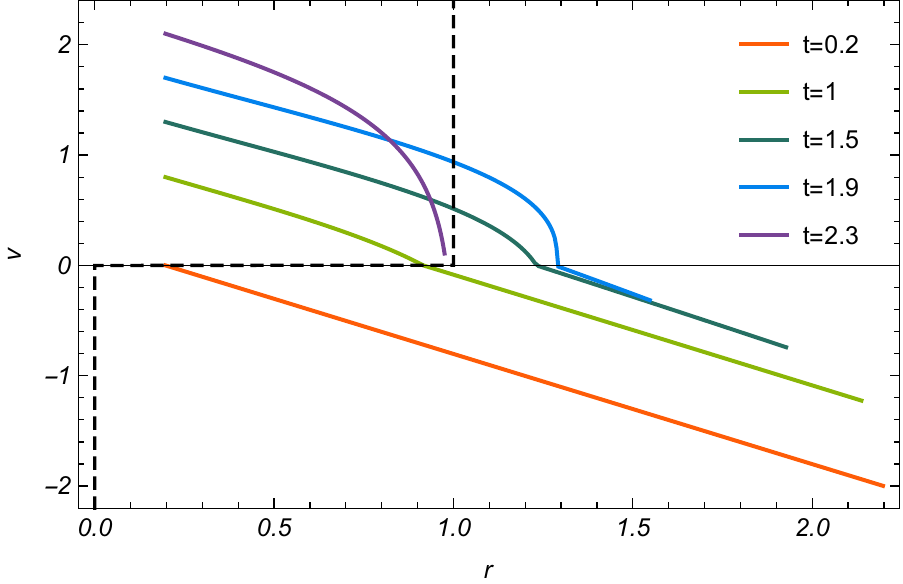}
\hspace*{0.8cm}
\includegraphics[scale=0.845]{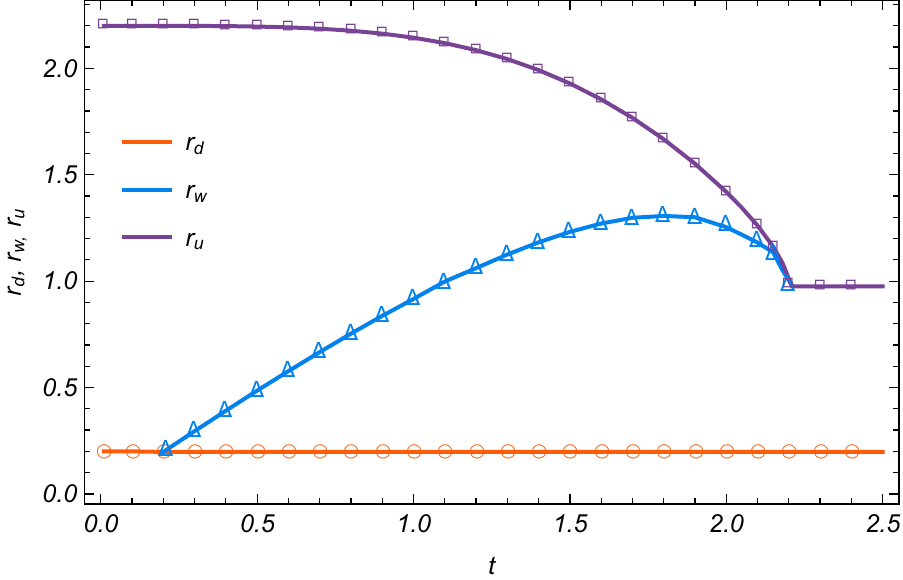}
\end{center}
\caption{\textit{Left}: The profile of $v(r)$ for fixed $h=0.4, \ell=2$ and different values of boundary time. The dashed line indicates the apparent horizon. \textit{Right}: Time evolution of $r_d, r_w$ and $r_u$. Clearly, $r_w$ always obeys the bound $r_d\leq r_w \leq r_u$. The markers show the corresponding numerical results. Here we have set $r_h=1$}
\label{fig:vranalytic}
\end{figure}

Now we are equipped with all we need to calculate the time dependence of EWCS analytically. Upon substituting the profiles of $v_a(r)$ and $v_b(r)$ into eq. \eqref{funcEWCS}, EWCS can be evaluated by separately evaluating the integral on the portion of the
geodesic above the shell and that below the shell as follows
\bea
E_W^{\rm AdS}&=&\frac{1}{4G_N}\int_{r_w}^{r_u} dr\frac{\sqrt{-2v_a'-{v_a'}^2}}{r}=\frac{1}{4G_N}\log\frac{r}{1+\sqrt{1+\mathcal{Q}_a^2r^2}}\Big|_{r_w}^{r_u},\\
E_W^{\rm BB}&=&\frac{1}{4G_N}\int_{r_d}^{r_w} dr\frac{\sqrt{-2v_b'-f_b(r){v_b'}^2}}{r}=\frac{1}{4G_N}\log\frac{r}{r_h^2+r_h\sqrt{r_h^2+(\mathcal{Q}_b^2-1)r^2}}\Big|_{r_d}^{r_w}.
\eea
The final result then becomes
\bea\label{finalew}
E_W=E_W^{\rm AdS}+E_W^{\rm BB}=\frac{1}{4G_N}\log\left(\frac{r_u}{r_d}\frac{1+\sqrt{1+\mathcal{Q}_a^2r_w^2}}{1+\sqrt{1+\mathcal{Q}_a^2r_u^2}}\frac{r_h^2+r_h\sqrt{r_h^2+(\mathcal{Q}_b^2-1)r_d^2}}{r_h^2+r_h\sqrt{r_h^2+(\mathcal{Q}_b^2-1)r_w^2}}\right).
\eea
In principle we should write the above result in terms of the boundary quantities such as $h$ and $2\ell+h$. Finding analytic solution of HRT turining points, i.e., $(v_t, r_t)$, in the terms of the width of entangling region is subtle. Here we recall that it was shown in \cite{Balasubramanian:2011ur} that the relation between $r_t$ and $X$ as a function of boundary time, i.e., $t$, can be expressed analytically in closed form as follows
\bea\label{X}
X=\frac{2r_h}{\rho}\frac{c}{s}+r_h\log\left(\frac{2(1+c)\rho^2+2s\rho-c}{2(1+c)\rho^2-2s\rho-c}\right),\;\;\;\;c=\sqrt{1-s^2},
\eea
with
\bea\label{rho}
\rho=\frac{1}{2}\coth\frac{t}{r_h}+\frac{1}{2}\sqrt{{\rm csch}^2\frac{t}{r_h}+\frac{1-c}{1+c}},
\eea
where $\rho=\frac{r_h}{r_c}$ and $s=\frac{r_c}{r_t}$. Note that $r_c$ is the value of $r$ at which $\Gamma_X$ intersects the null shell. Also  $v_t$ can be find as follows
\bea
v_t=r_c-r_t,
\eea
which is due to the matching condition for $\Gamma_X$\cite{Liu:2013qca}. With these expressions, the profile of the minimal geodesic and EWCS in eqs. \eqref{BBvr}, \eqref{adsvr} and \eqref{finalew} are implicitly expressed entirely in terms of boundary quantities, e.g., $h, \ell$ and $T$. In the left panel of Fig. \ref{fig:vranalytic}, the profile for EWCS which is determined from the analytic expressions in the $(v, r)$ plane are plotted for different values of boundary time. Further, we show the full evolution of $r_d, r_w$ and $r_u$ for a fixed $h$ and $\ell$ in the right panel of Fig. \ref{fig:vranalytic}. The markers in this figure show the numerical results which coincide with our analytical expressions in the thin shell limit.
We will explore some universal features in the time evolution of  EWCS using the closed form expression in detail in the next section.
\subsection{Regimes in the growth of EWCS}

The closed form expression for EWCS given by eq. \eqref{finalew} enables us to directly extract the different scaling behavior in various regimes during the thermalization process. In this section we will study these scaling regimes in more detail. In our setup, the main boundary quantities which may affect the behavior of holographic dual for EWCS during the thermalization process are $h, \ell$ and $T$. On the other hand, the dual geometric entities in the bulk which may govern the evolution of EWCS are $r_d, r_u$ and $r_h$. Also our numerical and analytical results in the previous sections suggest that the time dependence of $E_W$ could be associated to $r_w$, that ranges from $r_d$ (i.e., close to the turning point of $\Gamma_h$) at early times, to $r_u$ (i.e., close to the turning point of $\Gamma_{2\ell+h}$) at late times. In fact, $r_d$ and $r_u$ are also depend on time, so that $r_w$ is not a monotonically increasing function of time to ensure $r_d\leq r_w \leq r_u$ at all times. Based on these results, an immediate conclusion is that the evolution of EWCS is characterized by different scaling regimes depending on $r_w(t)$ which we will examine further in the following.

\subsubsection*{Early Growth}
At early times, the null shell does not reach $\Sigma$ which lies entirely in AdS geometry, and hence the EWCS is a fixed constant given by the vacuum value. The early growth of EWCS starts immediately after the shell intersects with $\Sigma$, i.e., $r_w\sim r_d$ and $v_d\sim v_{\rm shell}=0$. In other words, there exists a sharp time $t_1$ after which $\Gamma_h$ lies entirely in the black brane region and hence reduces to that in a static BTZ geometry. Indeed, $t_1$ is the saturation time for HEE corresponding to a boundary region with width $h$. Further, the boundary quantities $t$ and $h$ can be fixed in terms of bulk parameters, i.e., $v_d$ and $r_d$ using eqs. \eqref{Xbb} and \eqref{vtd3} as follows
\bea\label{ht}
h=r_h\log\frac{r_h+r_d}{r_h-r_d},\hspace*{1.5cm}t=v_d+\frac{r_h}{2}\log \frac{r_h+r_d}{r_h-r_d}.
\eea
Combining the above equations then yields
\bea
t=\frac{h}{2}+v_d>t_1.
\eea
On the other hand, because we consider $h\lesssim t \ll r_h\ll \ell$, $\Gamma_{2\ell+h}$ lies almost entirely in the AdS region and the point of intersection is very close to the boundary, i.e., $r_c\ll r_h$. 
From eqs. \eqref{X} and \eqref{rho}, we can expand $r_c$ and $r_u$ for early times to find
\bea\label{rcru}
r_c=t-\frac{t^3}{12r_h^2}+\cdots,\;\;\;\;\;r_u=\left(\ell+\frac{h}{2}\right)\left(1+\frac{t^4}{144r_h^4}+\cdots\right).
\eea
In the large $\ell$ limit we can approximate the value of $\mathcal{Q}_a, \mathcal{Q}_b$ and $r_w$ in eqs. \eqref{Qb}, \eqref{Qa} and \eqref{rweq} by that at $r_u\rightarrow \infty$. In this way, using $v_u=r_c-r_u$ we find
\bea\label{QAQB}
\mathcal{Q}_a=2\frac{r_c-r_w}{r_u^2}+\cdots,\;\;\;\;\;\;\;\;\;\mathcal{Q}_b=-\frac{r_w}{2r_h}+\frac{r_c-r_w}{r_h}\frac{2r_h^2-r^2_w}{r_u^2}+\cdots,
\eea
where $r_w$ is given by
\bea\label{rwearly}
r_w=2r_h\frac{r_h(p^2-1)+r_d(p^2+1)}{r_h(p+1)^2+r_d(p^2-1)},\;\;\;\;\;\;\;\;p=e^{\frac{v_d}{r_h}}.
\eea
The above solution for $r_w$, shows that in eqs. \eqref{Qb} and \eqref{Qa} one must pick the minus and plus sign for $\mathcal{Q}_b$ and $\mathcal{Q}_a$, respectively. To see this, note that when $\Gamma_h$ starts intersecting the null shell, i.e., $v_d\sim0$, we should have $r_w\sim r_d$ as satisfied by \eqref{rwearly}. Indeed choosing other signs this condition is not satisfied. 
Using the above relations for the early time limit (in which case, $t\ll r_h$ ), eq. \eqref{finalew} yields
\bea
E_W=\frac{1}{4G_N}\left(\log\frac{r_u}{r_d}+\frac{t^2}{4r_h^2}+\cdots\right).
\eea
We can use eqs. \eqref{ht} and \eqref{rcru} as well as $h\ll \ell$ condition, to rewrite it as follows
\bea\label{quadraticd2}
E_W=E_W^{\rm vac.}+\pi\mathcal{E}t^2+\cdots,
\eea
where $E_W^{\rm vac.}$ is the vacuum contribution given by eq. \eqref{ewd2vac} and $\mathcal{E}$ is the energy density given by
eq. \eqref{ESC} with $d=2$.
Therefore at early times, i.e., $h\lesssim t \ll r_h$,  EWCS grows quadratically and the rate of growth is a fixed constant proportional to the energy density. Similar scaling behavior was found for the early growth of entanglement entropy in \cite{Liu:2013qca}.

\subsubsection*{Linear growth}
As we discussed above, for $t>t_1$, $\Gamma_h$ lies on a constant time slice outside the horizon and is time independent. Hence, $r_d$ remains fixed and we can use eq. \eqref{ht} to find $r_d$ and $v_d$. 
One might note that in fact the time evolution of $\Sigma$ is then largely governed by properties of $\Gamma_{2\ell+h}$ which go through both
AdS and black brane regions. Based on our numerical results we  expect that this regime corresponds to $h\ll r_h\ll t \ll \ell$.
In this regime we can expand eqs. \eqref{X} and \eqref{rho} to find
\bea
&&\frac{r_h}{r_c}=\frac{1}{2}+\frac{r_c}{4r_u}+\frac{2r_u}{r_c}e^{-2\frac{t}{r_h}}+\cdots,\nonumber\\
&&\ell+\frac{h}{2}=2\frac{r_h}{r_c}r_u+t+r_h\log\frac{r_c}{r_u}+\cdots.
\eea
Solving these at leading order then yields
\bea
r_u=\ell+\frac{h}{2}-t+\cdots,\hspace*{2cm}r_c=2r_h+\cdots.
\eea
Once again in the $r_u\rightarrow \infty$ limit we can use eqs. \eqref{QAQB} and \eqref{rwearly} for $\mathcal{Q}_a$, $\mathcal{Q}_b$ and $r_w$, noting that in this case we should consider the above expressions for $r_u$ and $r_c$. 
Upon substituting these results into eq. \eqref{finalew} and expand for large $\ell$, the resulting EWCS is then
\bea
E_W=E_W^{\rm vac.}+\frac{1}{4G_N}\frac{t}{r_h}-\frac{1}{4G_N}\log 4+\cdots,
\eea
which can be rewrite as follows
\bea\label{linear2d}
E_W=E_W^{\rm vac.}+s_{\rm eq.}t-\frac{c}{6}\log 4+\cdots,
\eea
where $s_{\rm eq.}$ is the equilibrium thermal entropy density given by eq. \eqref{ESC}. 
It is worth to mention that the above constant rate precisely matches with the previous results of 
\cite{Moosa:2020vcs,2001.05501}. Similar scaling behavior was found during intermediate stages of time evolution of entanglement entropy in \cite{Liu:2013qca}.
\subsubsection*{Saturation}
At late times, the tip of $\Gamma_{2\ell+h}$ approaches the null shell, i.e., $r_w\rightarrow r_u$ from below and $v_u\rightarrow 0$. Thus, we can expand the relevant quantities in small $r_u-r_w$ and $v_u$. Indeed, in this case $\Sigma$ lies entirely in the black brane region and hence reduces to that in a static BTZ geometry. Recall that an essential assumption in evaluating EWCS is that both HRT hypersurfaces, i.e., $\Gamma_{h}$ and $\Gamma_{2\ell+h}$, correspond to the same boundary time $t$. We use this condition to simplify the calculation since for $v_u\sim 0$, $v_d$ can be expressed in terms of $r_d$ and $r_u$ as follows
\bea
v_d=v_u-\frac{r_h}{2}\log\frac{r_h-r_u}{r_h+r_u}+\frac{r_h}{2}\log\frac{r_h-r_d}{r_h+r_d},
\eea
where we have used eq. \eqref{vtd3}. Inserting the above expression in eqs. \eqref{Qb} and \eqref{rweq} and simplify the resultant equations yields 
\bea
\mathcal{Q}_a=\frac{r_w}{2r_h\sqrt{r_h^2-r_w^2}},\hspace*{1cm}\mathcal{Q}_b=0.
\eea
Combining the above relation with eq. \eqref{Qa} and expand for small $v_u$, we have  
\bea\label{ruvu}
r_w=r_u+\left(1-\frac{r_u^2}{2r_h^2}\right)v_u+\cdots. 
\eea
Upon substituting these results into eq. \eqref{finalew}, one finds that at leading order the resulting EWCS is then given by the same expression as in \eqref{ewd2bb}. Also note that according to eq. \eqref{ruvu} the $v_u\rightarrow 0$ limit coincides with $r_w\rightarrow r_u$ as expected.

Producing the different scaling behavior of EWCS shows that the results based on our analytic treatment are consistent with the previouslly numerical results.

\section{EWCS in Vaidya backgrounds: higher dimensions }\label{highdim}
In this section we generalise our studies to higher dimensional cases in specific directions. We will mainly focus on three dimensional boundary theory, because the interesting qualitative features of the thermalization process are independent of the dimensionality of the QFT. In order to investigate the behavior of EWCS during the thermalization process, we
consider two different types of global quench: a thermal quench and an electromagnetic quench. Once again, we consider subsystems consisting of equal width intervals as depicted in figure \ref{fig:regions}. For simplicity, we set $r_h=1$ and work with the rescaled quantity $\tilde{E}_W=\frac{4G_N}{\tilde{\ell}^{d-2}} E_W$ throughout the following. 

\subsection{Evolution after a thermal quench $(q=0)$}

Let us begin with the case of a thermal quench where the dual gravitional geometry is given by eqs. \eqref{staticmetric} and \eqref{staticmetric1} with $q=0$. In the following, we evaluate $E_W(t)$, defined in eq. \eqref{funcEWCS}, numerically for several values of $h, \ell$ and $T$. To do that we should solve eqs. \eqref{EOMHEE} and \eqref{EOMEWCS} to find the corresponding profiles for $\Gamma_h, \Gamma_{2\ell+h}$ and $\Sigma$. 
Below in figures \ref{fig:shmiew1} and \ref{fig:shmiew2}, we show the time evolution of HMI and EWCS for various $h$ and $\ell$.\footnote{The results for HMI with the same parameters was previously reported in \cite{Alishahiha:2014jxa}.} We also present the competition between connected and disconnected configurations in these figures for the same boundary regions. From these plots, one can infer that the qualitative features of the evolution is similar to the three dimensional case, as we explain below.

In figure \ref{fig:shmiew1}, we show various boundary quantities for the case of $hT < \ell T <1$. In the left
panel, we compare the values of $S_{\rm con.}(t)$ and $S_{\rm dis.}(t)$ to see which configuration is minimal during
the evolution. Clearly, in this case the connected configuration is always favored for any time. The middle panel demonstrates the evolution of HMI which is always nonzero and the right panel shows $E_W(t)$. At early times, i.e., $t \ll h$, the EWCS starts at the same value of the
AdS geometry given by eq. \eqref{ewt0}, then at $t \sim \mathcal{O}(h)$ quickly deviates from the vacuum value and
approaches a regime of linear growth. Once again, we observe a period of time during which the growth of the EWCS is quadratic. The regime with linear growth persists all the way up to $t\sim \mathcal{O}(\ell+h)$ where $E_W$ reaches its maximum value. At late times, $E_W$ decreases and saturates to the equilibrium value corresponding to the black brane geometry given by eq. \eqref{ewBB}.
\begin{figure}
\begin{center}
\includegraphics[scale=0.6]{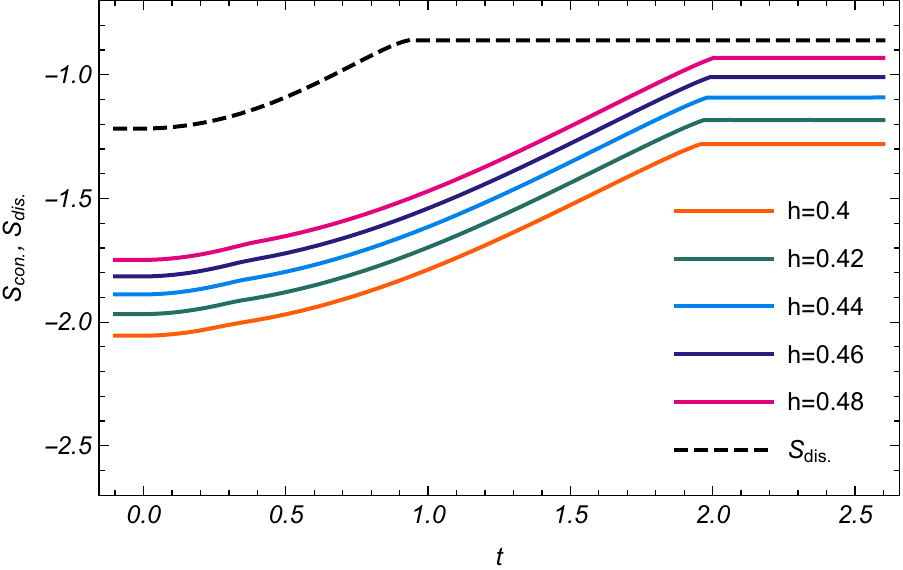}
\hspace*{0.05cm}
\includegraphics[scale=0.41]{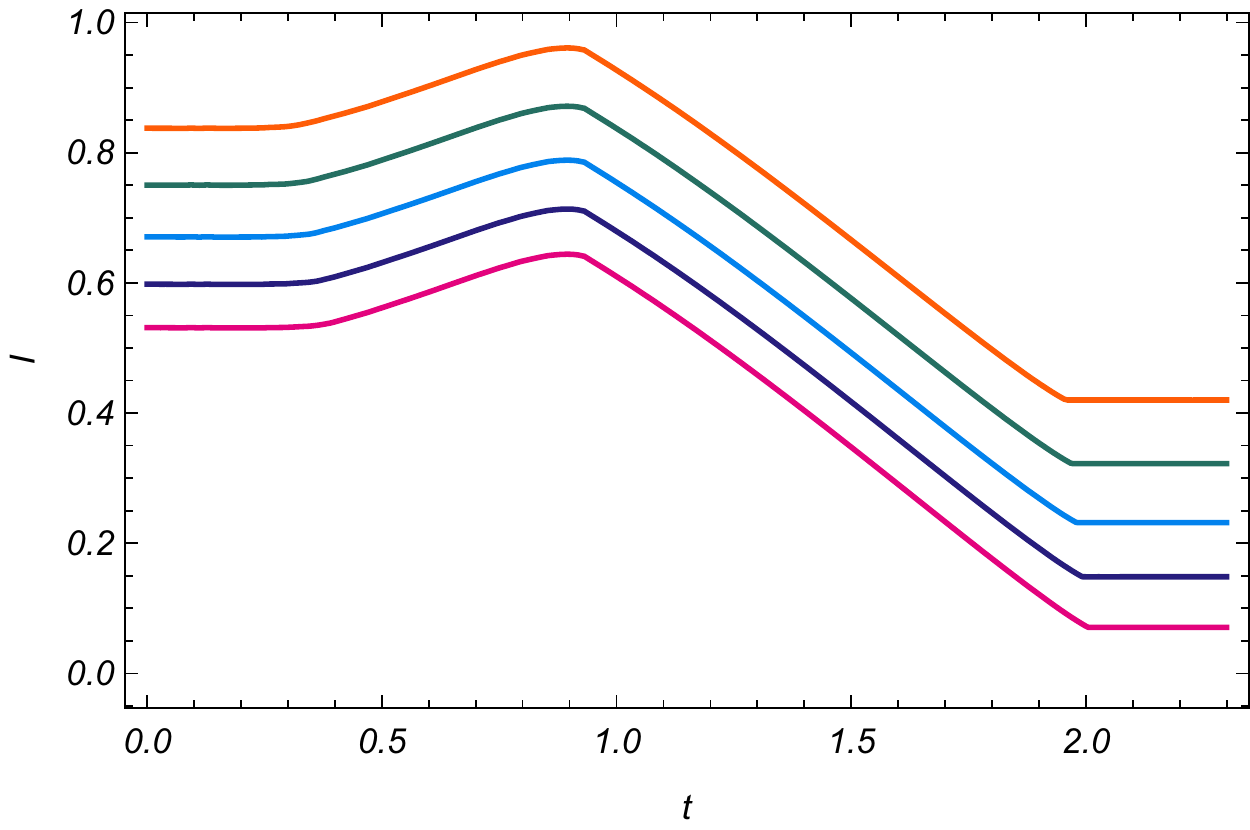}
\hspace*{0.05cm}
\includegraphics[scale=0.41]{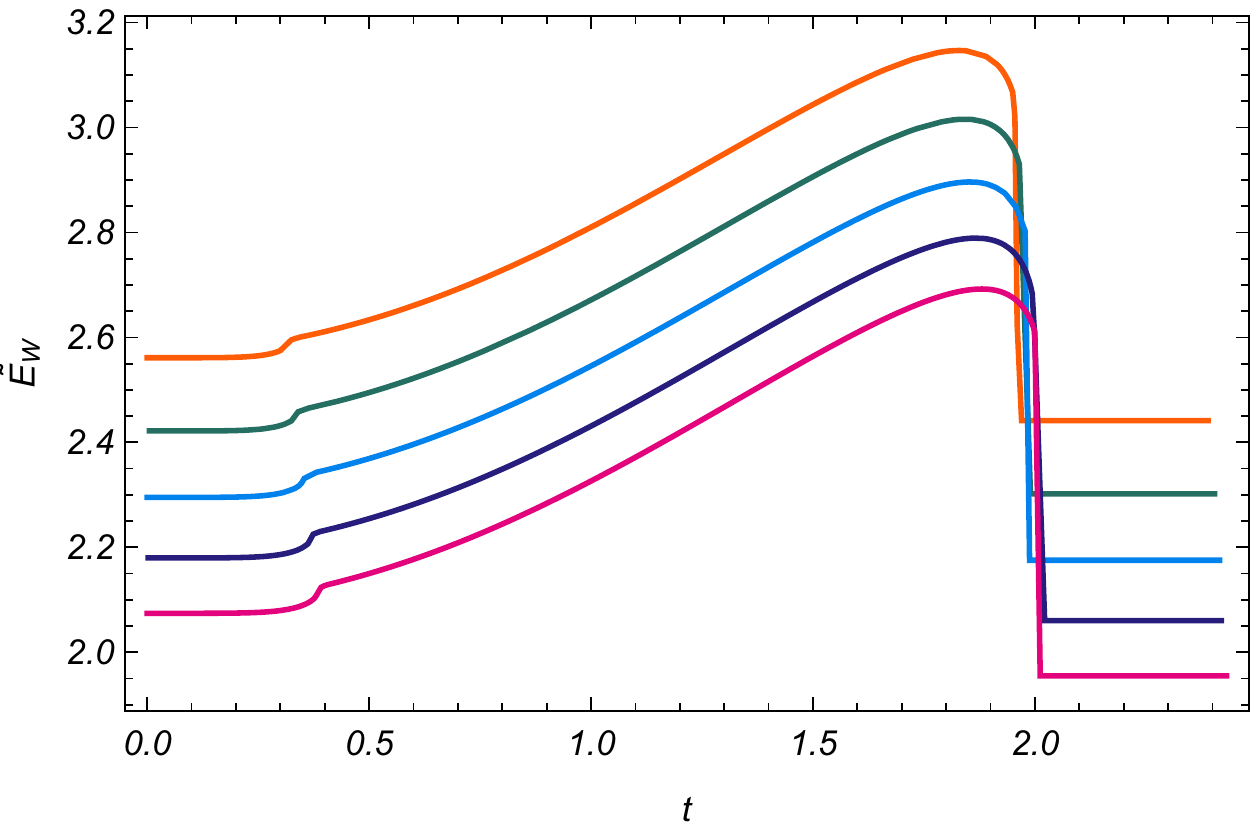}
\end{center}
\caption{\textit{Left}: The HEE for connected and disconnected configuration for different values of the separation between subregions. The connected configuration is always favored for any boundary time. \textit{Middle}: The HMI as a function of time which is positive during the entire evolution. \textit{Right}: The EWCS as a function of boundary time which saturates to a finite value. Here we have set $\ell=1.18$ and $d=3$.}
\label{fig:shmiew1}
\end{figure}

In figure \ref{fig:shmiew2} we demonstrate the same boundary quantities for the case of $hT < 1< \ell T$. According to the left panel, although the connected configuration has the minimal area at early time, the late time behavior is governed by the disconnected configuration. The critical time when this transition happens is approximately $t\sim \mathcal{O}(\ell-h)$. The main difference with the previous case is the late time behavior where EWCS displays a discontinuous transition and immediately saturates to zero. 

Comparing EWCS in the above two examples, we find that for small entangling regions, it transitions between two different constant regimes corresponding to vacuum and thermal values, without much of a linear regime in between. Further, as the width of the entangling region becomes larger, the region with linear dependence becomes more pronounced. If we fit the curves in the right panel of figure \ref{fig:shmiew2} in the linear growth regime, we find $\Delta \tilde{E}_W\sim v_w t$ where the best fit gives $v_w\approx 0.68$. Note that, the slope seems more or less the same independent of $h$.

\begin{figure}
\begin{center}
\includegraphics[scale=0.58]{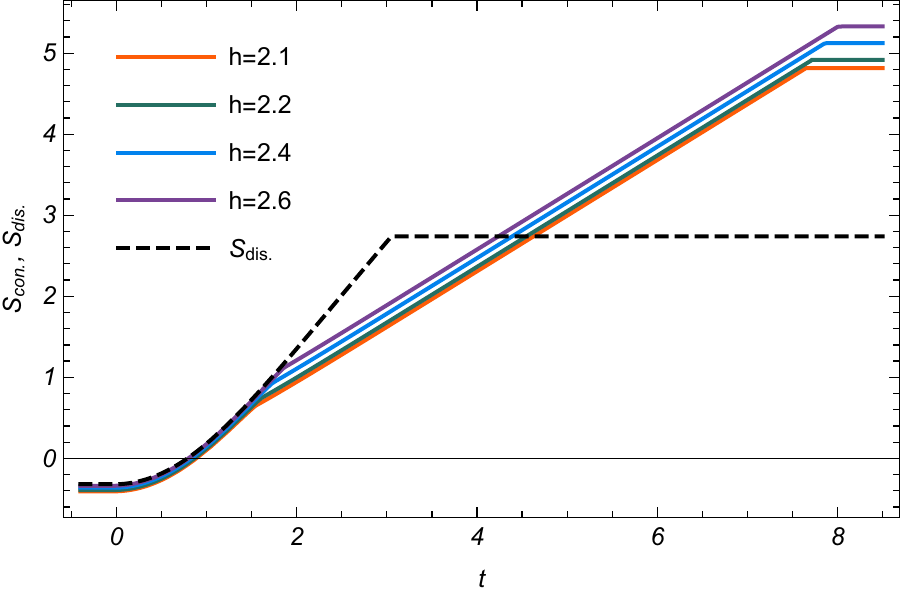}
\hspace*{0.05cm}
\includegraphics[scale=0.415]{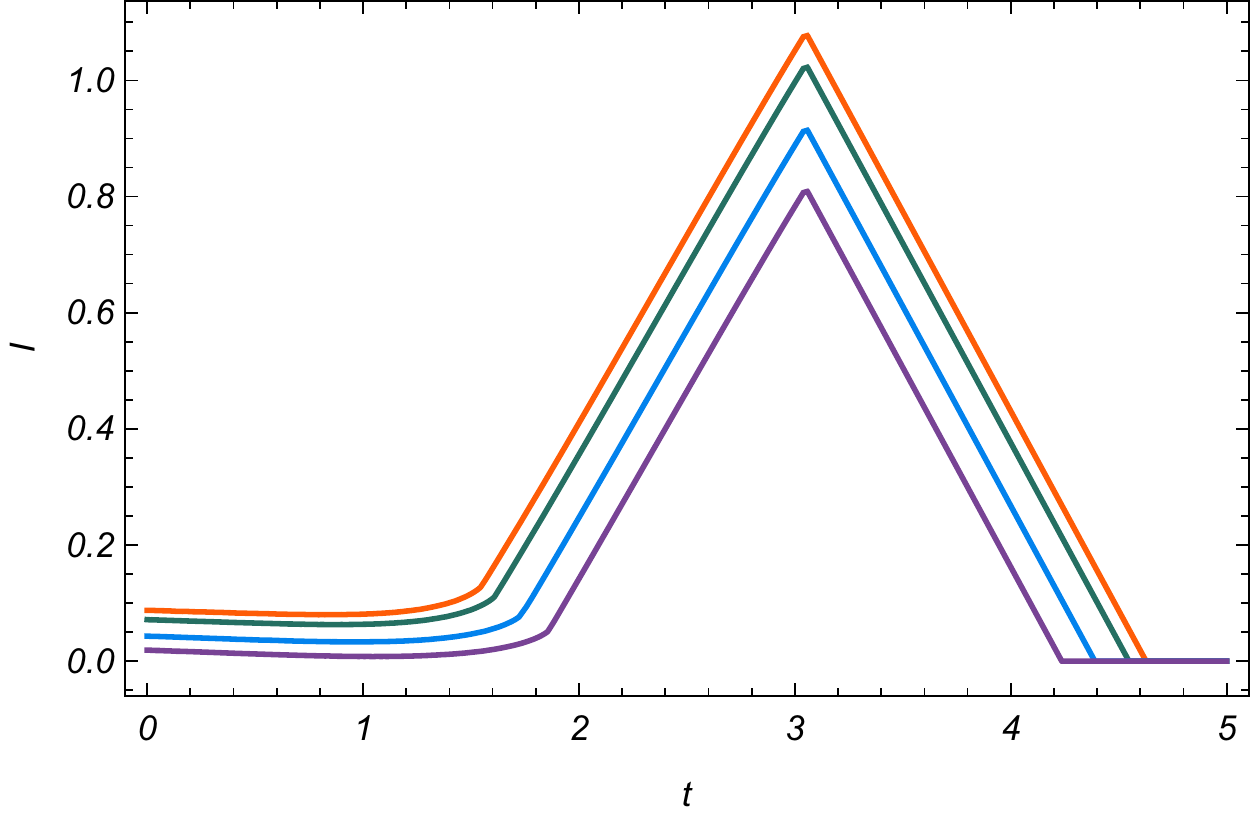}
\hspace*{0.05cm}
\includegraphics[scale=0.415]{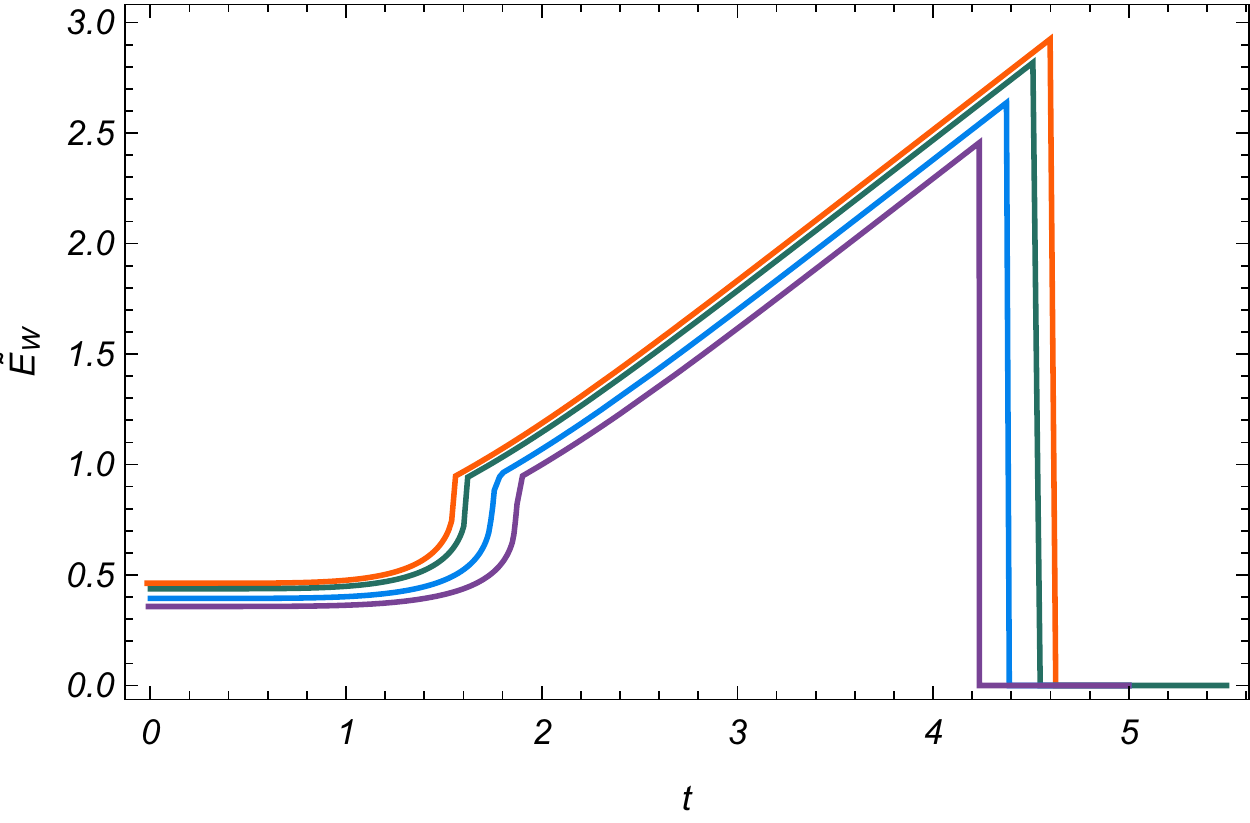}
\end{center}
\caption{\textit{Left}: The HEE for connected and disconnected configuration for different values of $h$. At late times, the disconnected configuration is favored. \textit{Middle}: The HMI as a function of time which vanishes at late times. \textit{Right}: The EWCS as a function of boundary time which saturates to zero. Here we have set $\ell=4.5$ and $d=3$.}
\label{fig:shmiew2}
\end{figure}
\subsection{Evolution after a electromagnetic quench $(q\neq 0)$}

In this section, we study the case of an electromagnetic quench where a thin shell of charged null fluid collapsing in empty AdS
to form a charged black brane. In the following, we consider two specific setups in which we evaluate the behavior of $E_W(t)$. We start by considering the case of a general electromagnetic quench where the final equilibrium geometry is a charged black brane at finite temperature and study scaling of EWCS during the thermalization. We then choose the system to be entirely non-thermal by approaching the extremal black brane solution whose blackening factor at late times is given by eq. \eqref{fext}.

\subsubsection*{Thermal electromagnetic quench $(T\neq 0)$}
In this case we should solve eq. \eqref{EOMEWCS} where the blackening is given by eq.\eqref{vaidyametric}. Before we perform the
computation, let us begin by discussing a new feature which may arise in this model. Indeed, it was demonstrated in \cite{Caceres:2013dma} that this model generically violates the NEC in some region in the bulk, although in some special
cases the HRT surfaces that compute HEE do not probe this region. When the extremal surfaces do probe this region the dual QFT would violate SSA which is problematic. To avoid this, we choose the time-dependent mass and charge functions such that the corresponding background respects SSA. This condition still leaves us with an enormous freedom in choosing these functions. In what Follows, we will focus on a simple choice where the mass and charge functions are given by
\bea
m(v)=\frac{1}{2}\left(1+\tanh\left(\frac{v}{0.01}\right)\right),\;\;\;\;\;\;\;\;q(v)=0.9\;m(v)^{\frac{2}{3}}.
\eea
In \cite{Caceres:2013dma} it was shown that considering the above choice the HRT surfaces attached to the
boundary never cross into that \textit{forbidden region} and SSA is satisfied. 
\begin{figure}
\begin{center}
\includegraphics[scale=0.417]{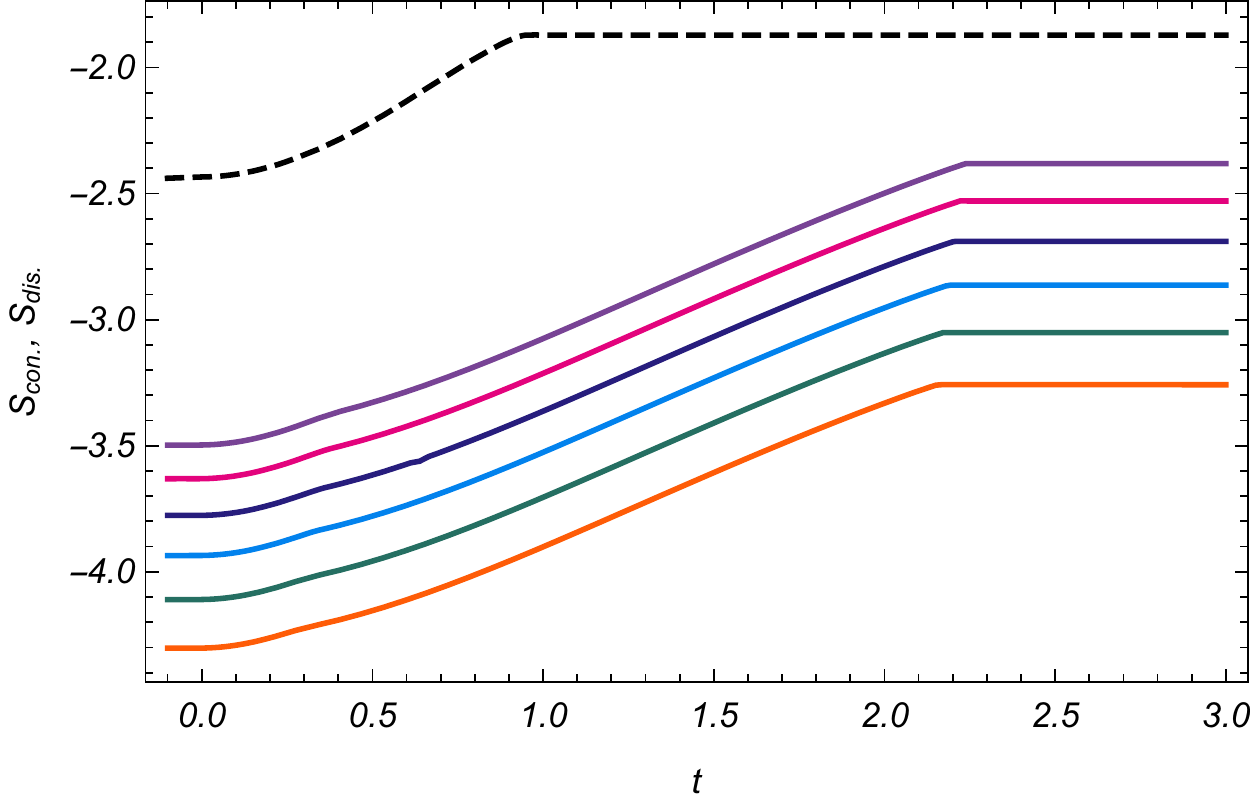}
\hspace*{0.05cm}
\includegraphics[scale=0.589]{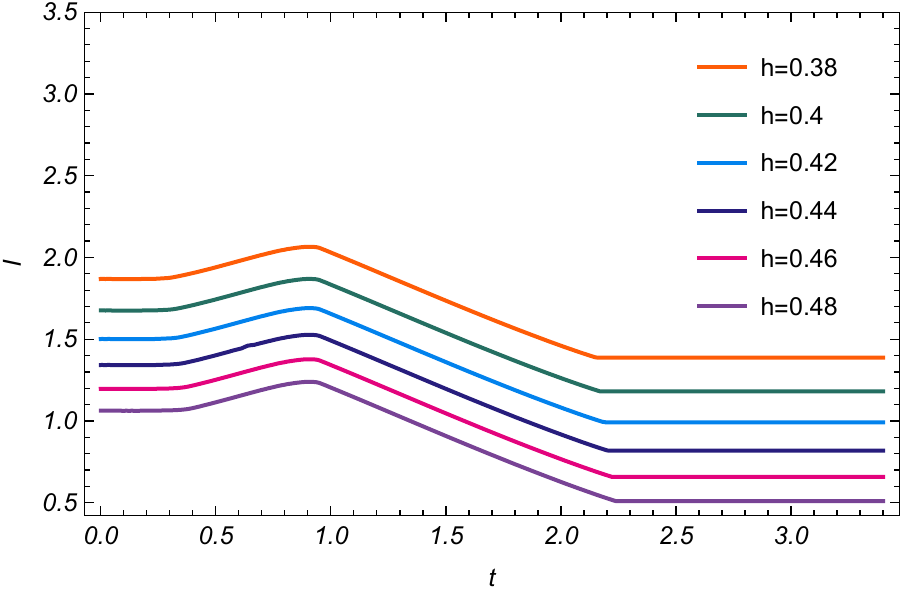}
\hspace*{0.05cm}
\includegraphics[scale=0.42]{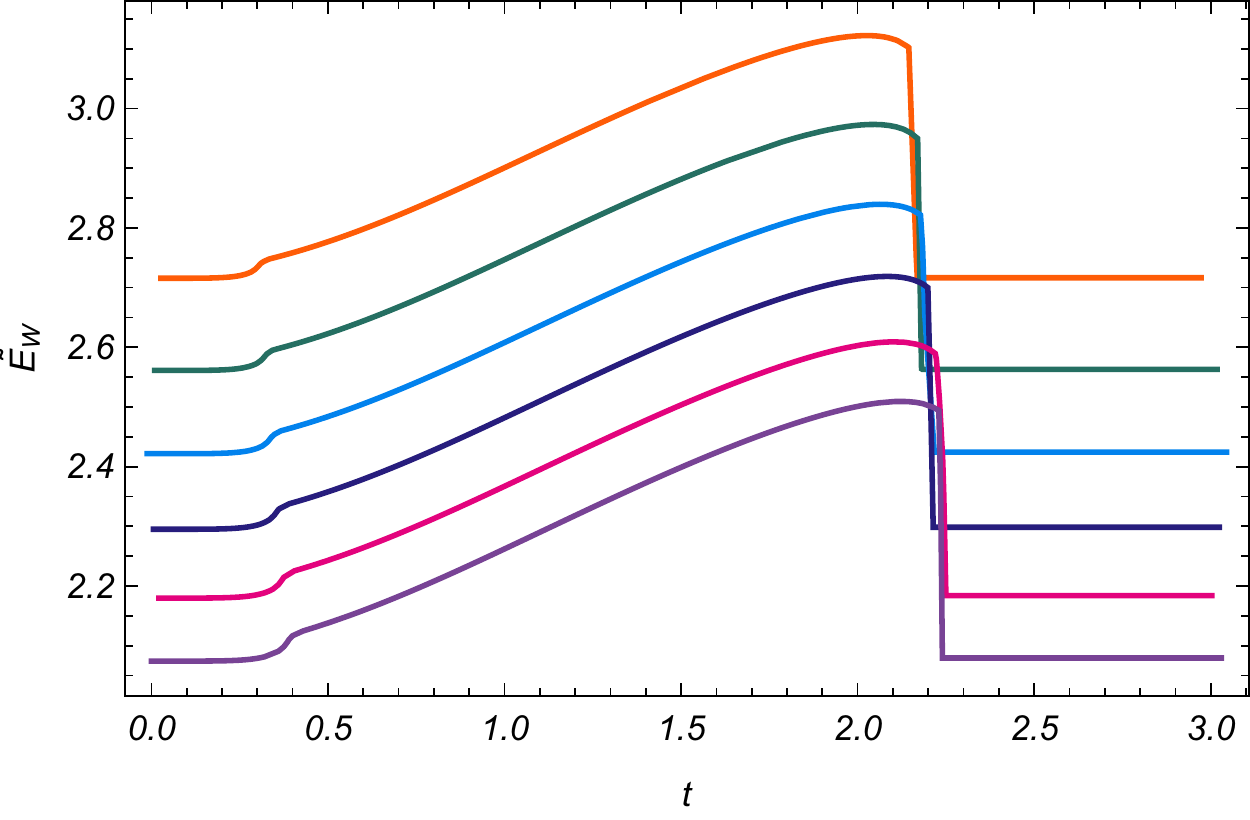}
\end{center}
\caption{The HEE (left), HMI (middle) and EWCS (right) as functions of boundary time for different values of the separation between subregions in a charged black brane geometry. The connected configuration is always favored for any boundary time and EWCS saturates to a finite value. Here we have set $\ell=1.18$ and $d=3$.}
\label{fig:charged-1}
\end{figure}

\begin{figure}
\begin{center}
\includegraphics[scale=0.411]{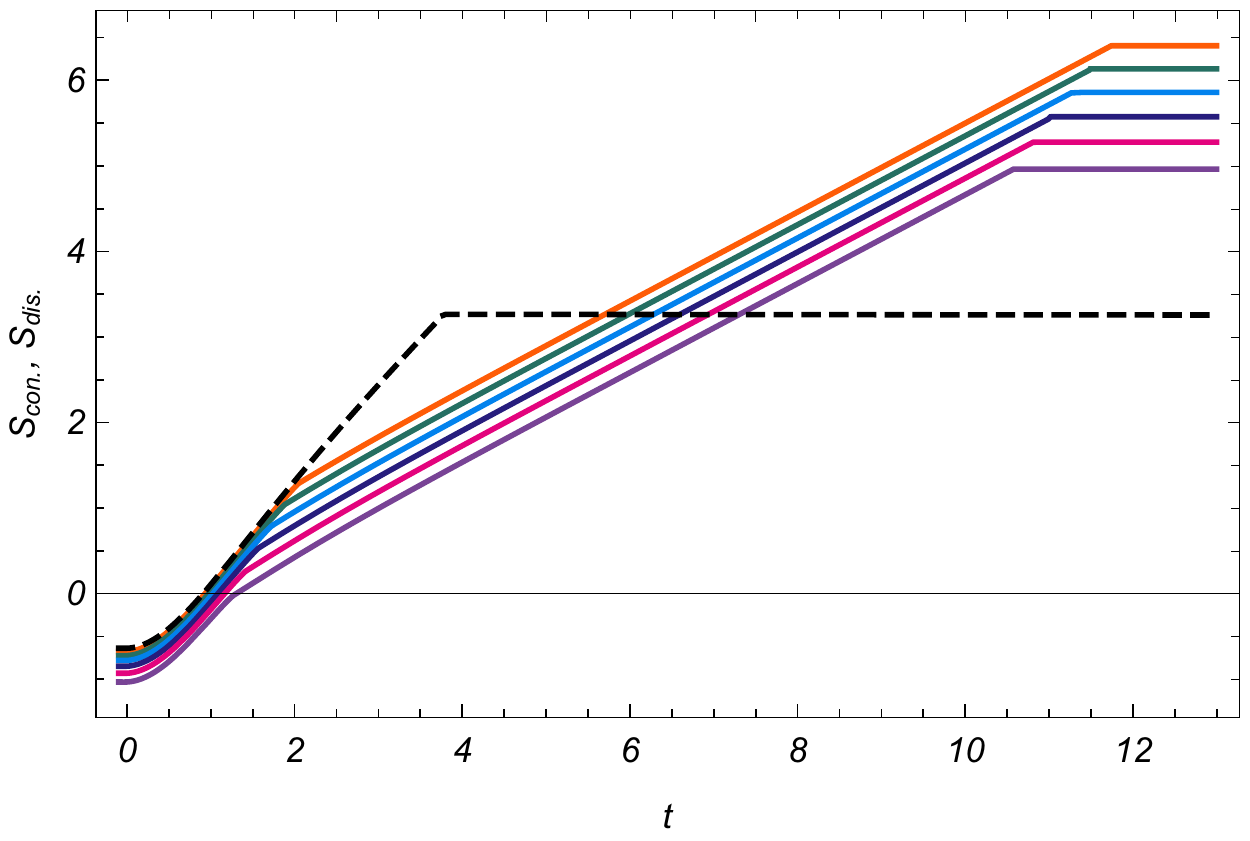}
\hspace*{0.05cm}
\includegraphics[scale=0.417]{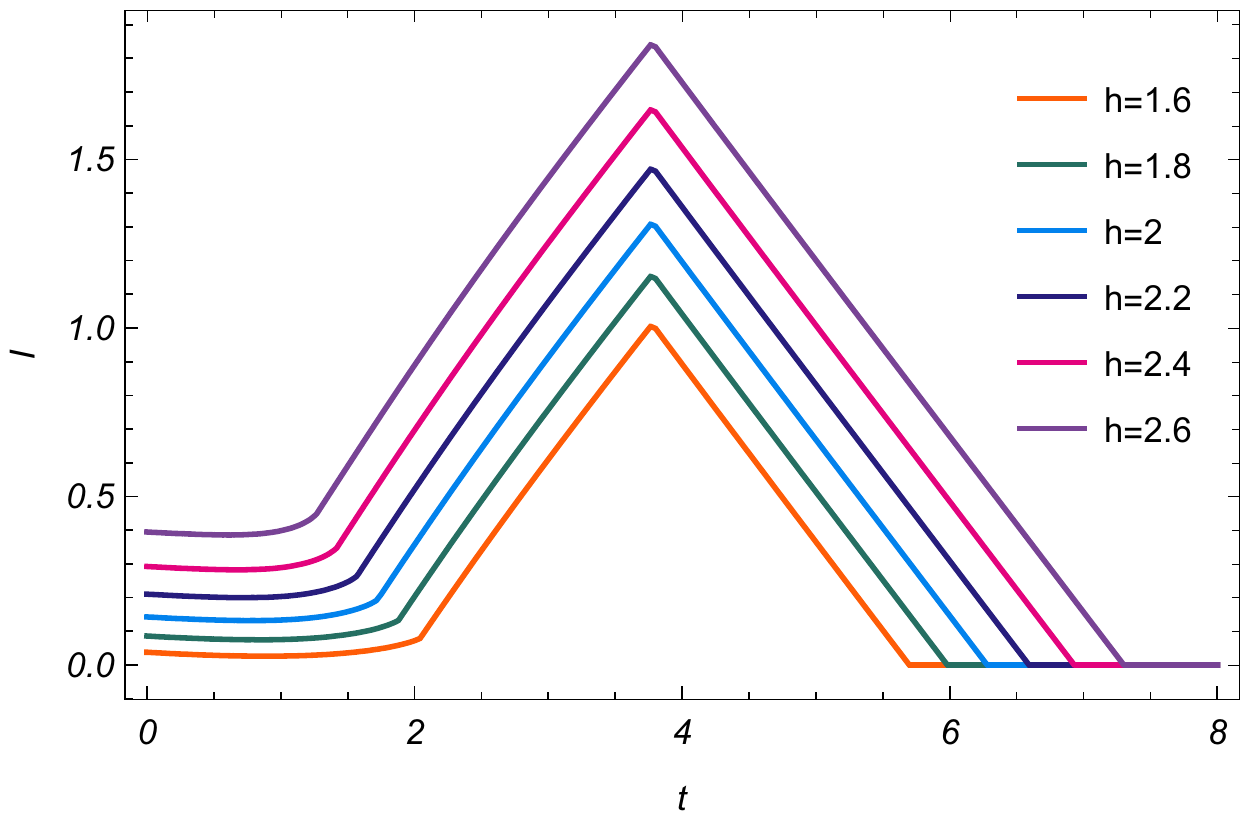}
\hspace*{0.05cm}
\includegraphics[scale=0.415]{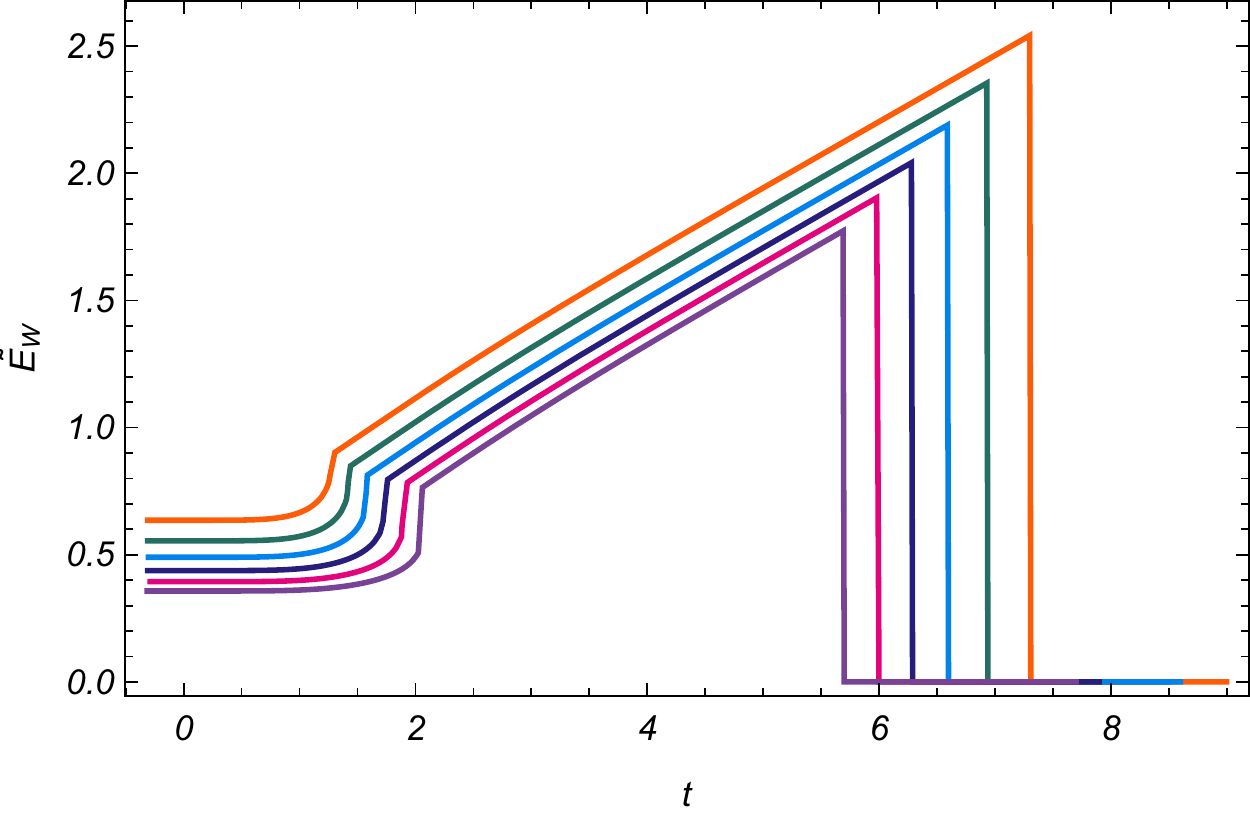}
\end{center}
\caption{The HEE (left), HMI (middle) and EWCS (right) as functions of boundary time for different values of the separation between subregions in a charged black brane geometry. At late times, the disconnected configuration is favored and EWCS saturates to zero. Here we have set $\ell=4.5$ and $d=3$.}
\label{fig:charged-2}
\end{figure}

In figures \ref{fig:charged-1} and \ref{fig:charged-2}, we plot the results for the HEE, HMI and EWCS for various
values of the parameters. Once again, one can see that in the period of linear growth, the slope seems more or less the same independent of the separation between subregions. Also note that when the disconnected configuration is favored at late times the saturation time of EWCS is smaller than the saturation time of the HEE associated with $\Gamma_{2\ell+h}$ as expected (see figure \ref{fig:charged-2}).

\begin{figure}
\begin{center}
\includegraphics[scale=0.41]{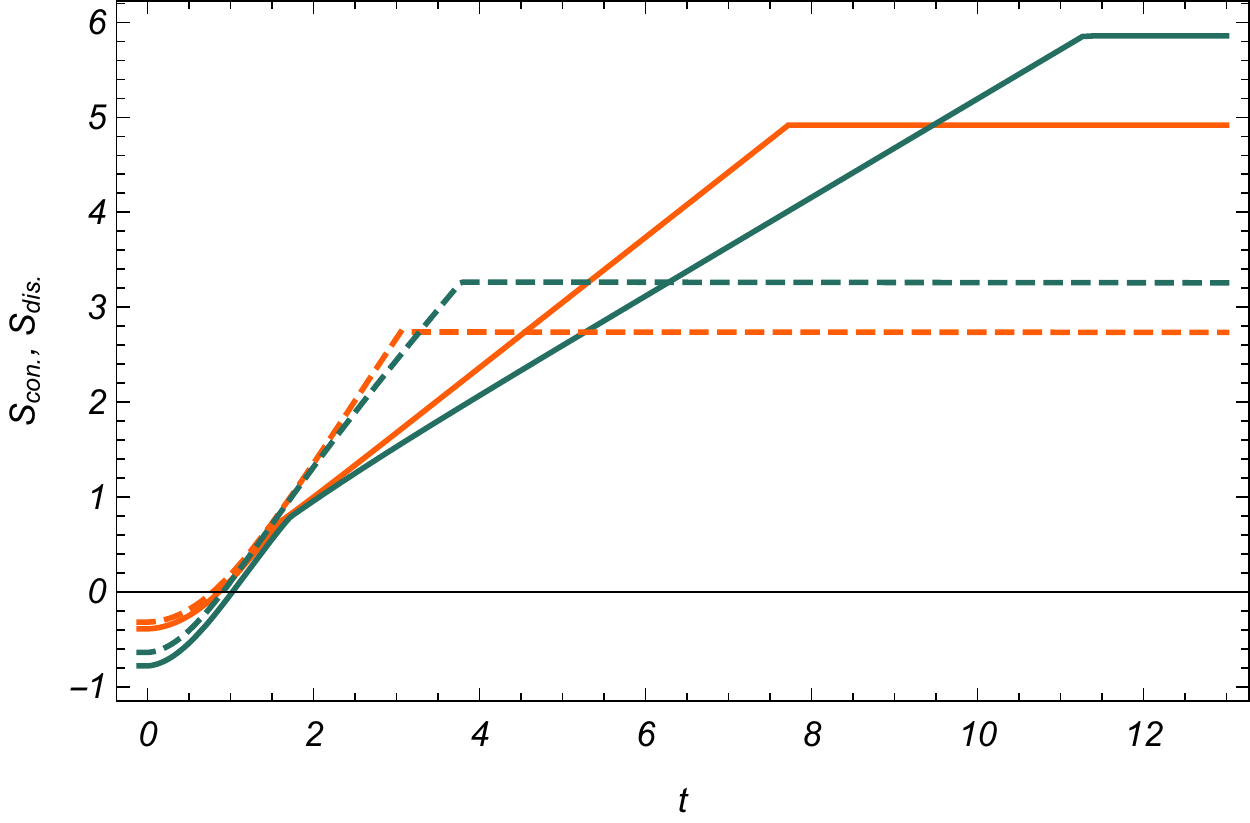}
\hspace*{0.05cm}
\includegraphics[scale=0.58]{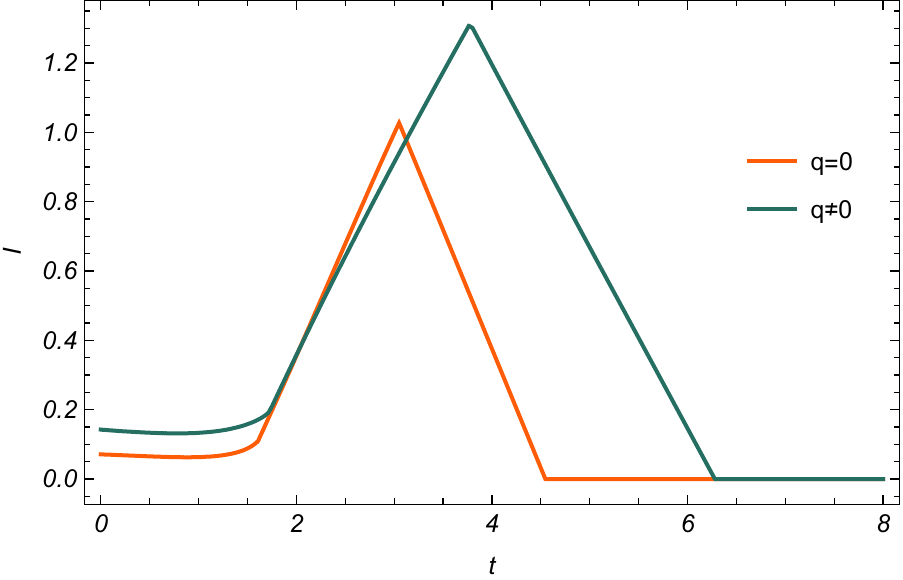}
\hspace*{0.05cm}
\includegraphics[scale=0.414]{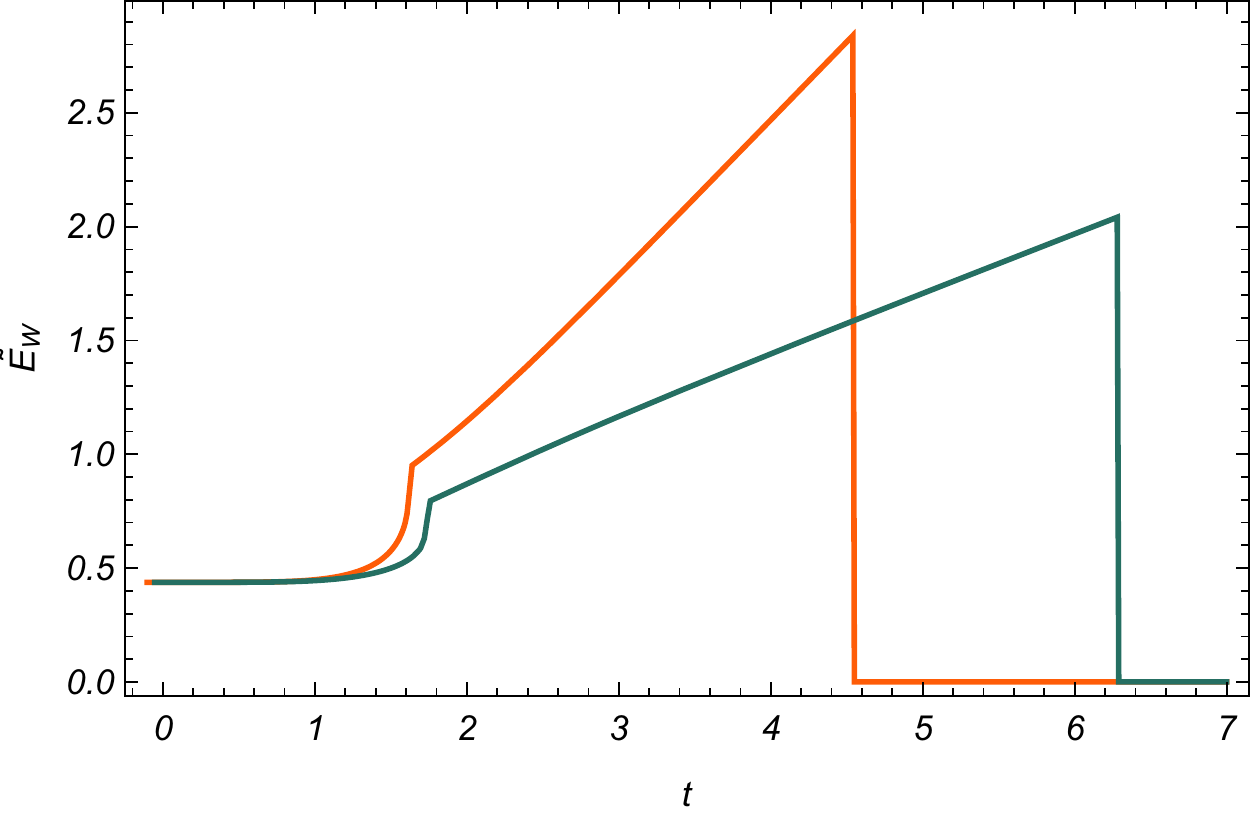}
\end{center}
\caption{Comparison of the time evolution of the HEE (left), HMI (middle) and EWCS (right) for neutral and charged quenches. Clearly when there is charge density, the equilibration process becomes less efficient. Here we have set $h=2.2, \ell=4.5$.}
\label{fig:compare}
\end{figure}

Note that in general, the qualitative features of the evolution is similar to the previous case with $q=0$. Let us emphasize that from the bulk perspective, it is natural to expect that turning on charged matter fields will slow down the thermalization process. On the other hand, from the boundary perspective, the equilibration process becomes less efficient due to the presence of charge density (or chemical potential) and one finds $t_s^{q=0}<t_s^{q\neq 0}$. The examples depicted in figure \ref{fig:compare} exhibit this behavior. To close this discussion, let us emphasize that, it was shown in \cite{Caceres:2012em} that in the case of AdS-RN-Vaidya the saturation of HEE is non-monotonous with respect to the chemical potential. We expect that similar non-monotonic behavior happens in the case of EWCS. The intuitive argument for this expectation is the dependence of EWCS on the configuration of (RT) HRT hypersurfaces which also fix the HEE. Indeed, regarding the results illustrated in figure \ref{fig:compare}, the dependence of saturation time on the chemical potential seems to be a universal behavior, independent of the corresponding entanglement measure one may choose. The detailed exploration of this behavior is outside the scope of this work and we leave it for future study \cite{WORK}.

\subsubsection*{Extremal electromagnetic quench $(T=0)$}
In the case of extremal electromagnetic quench where the system is entirely non-thermal, the blackening factor at late times is given by eq. \eqref{fext} corresponds to that of an extremal solution and we have
\bea\label{rhvextremal}
r_h^{-d}(v)=\frac{1}{2r_h^{d}} \left(1+\tanh\left(\frac{v}{v_0}\right)\right).
\eea
In this case, $m(v)$ and $q(v)$ are not independent and we consider eq. \eqref{rhvextremal} as the time-dependent profile for the horizon radius. Before we proceed, let us recall that for an extremal geometry the event horizon has a double zero. This feature plays an important role in the evolution of EWCS after an electromagnetic quench as we detail below. 

\begin{figure}
\begin{center}
\includegraphics[scale=0.59]{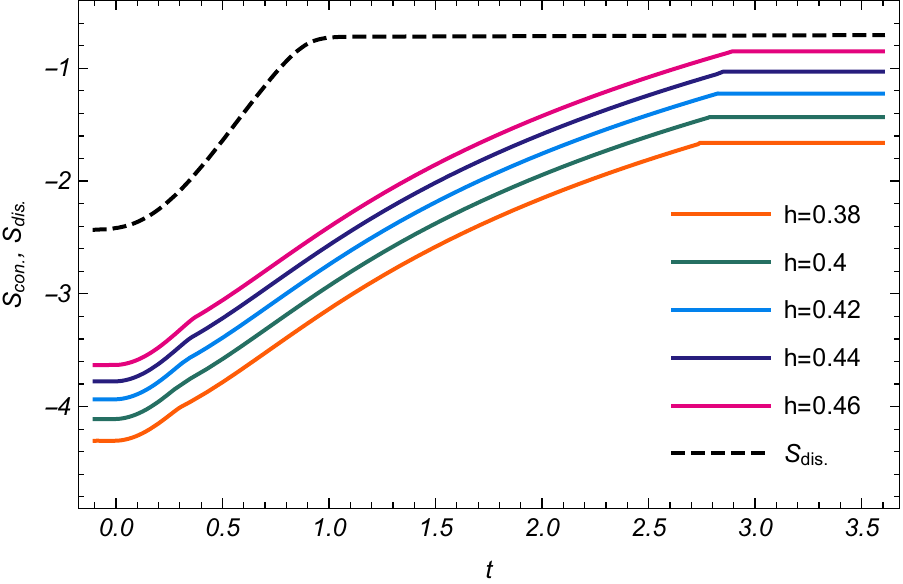}
\hspace*{0.05cm}
\includegraphics[scale=0.415]{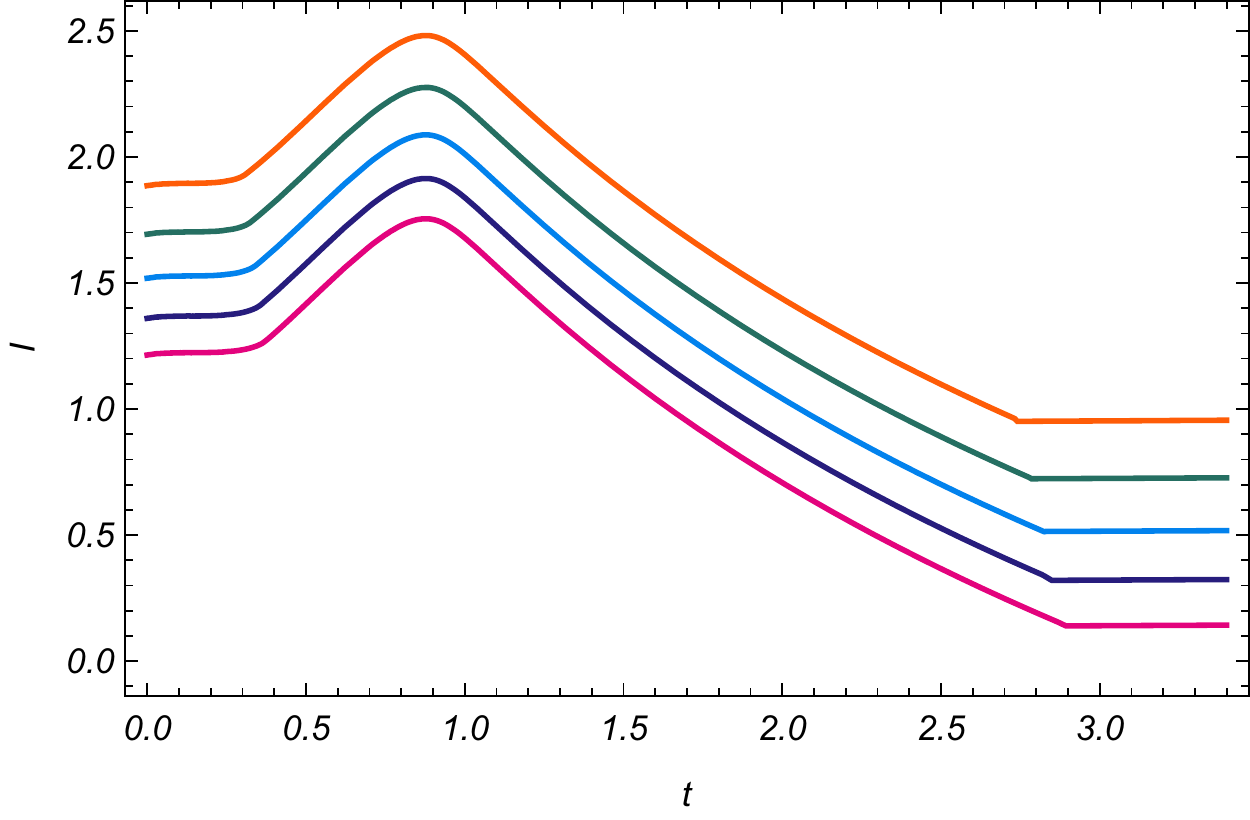}
\hspace*{0.05cm}
\includegraphics[scale=0.415]{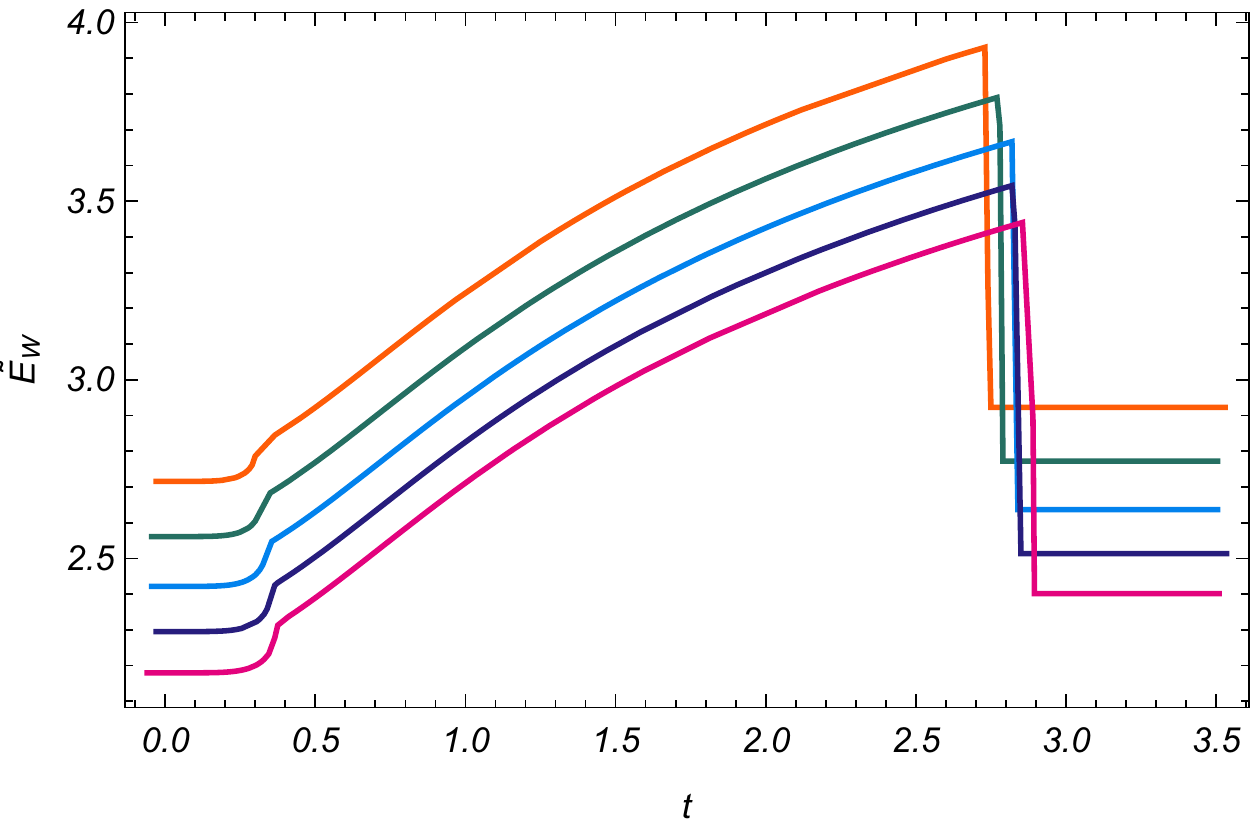}
\end{center}
\caption{The HEE (left), HMI (middle) and EWCS (right) as functions of boundary time for different values of the separation between subregions in an extremal black brane geometry. The connected configuration is always favored for any boundary time and EWCS saturates to a finite value. Here we have set $\ell=1.18, v_0=0.001$ and $d=3$.}
\label{fig:extremal-1}
\end{figure}

In figure \ref{fig:extremal-1}, we show the numerical results for a fixed $\ell$ and several values of $h$. In this case the connected configuration is always favored for any boundary time and EWCS saturates to a finite value. 
There is some interesting differences when comparing the behavior of the evolution after a electromagnetic quench here to the thermal quench in the previous section. For extremal cases, the regime of linear growth is replaced by a logarithmic growth. 
This behavior is inherited from the logarithmic scaling in the static extremal geometries as previously discussed in \cite{Albash:2010mv}, which has its origin in the double zero at the horizon.

\section{Conclusions and Discussions}\label{dis}
In this paper, we explored the time evolution of entanglement wedge cross section after a global quantum quench for a strip entangling region in
various geometries. We considered subsystems consisting of equal width intervals as depicted in figure \ref{fig:regions}. First, we focused on the simple case of $d=2$ and consider a thermal quench in detail where the final equilibrium state is dual to a BTZ geometry. In this case, much of the analysis could be carried out analytically. We have also extended these studies to $3+1$ dimensions, where we considered two different types of global quench: a thermal quench and an electromagnetic quench. In the following, we would like to summarize our main results and also discuss some further problems.
\begin{itemize}
\item  
In a $(2+1)$-dimensional bulk geometry, we found that the time evolution of EWCS is characterized by three different scaling regimes: 
an early time quadratic growth, an intermediate linear growth and a late time saturation. The main behavior in the evolution depends on $\Gamma_{2\ell+h}$ while $\Gamma_{h}$ is fixed and do not influence the time dependence of EWCS. 
To confirm these behaviors, we provided a numerical analysis and examined the various regimes in the
growth of EWCS in the thin shell limit. We have also found an analytically closed form expression for EWCS, which enables us to directly extract its scaling behavior in various regimes. Our results here show that at early times, i.e., $t \ll h$, the EWCS starts at the same value of the pure AdS geometry, then at $t \sim \mathcal{O}(h)$ grows quadratically and approaches a regime of linear growth. We found that  as the width of the entangling region becomes larger, the region with linear dependence becomes more pronounced. 
In analogy to the analysis in \cite{Liu:2013qca} and motivated by this linear growth we introduce a dimensionless rate of growth as follows
\bea
\mathcal{R}_W(t)\equiv\frac{1}{s_{\rm eq.}\tilde{\ell}^{d-1}}\frac{dE_W}{dt}.
\eea
Using eqs. \eqref{quadraticd2} and \eqref{linear2d} we find that for a BTZ geometry
\bea
\mathcal{R}_W(t)=\Bigg\{ \begin{array}{rcl}
&2\pi\frac{\mathcal{E}}{s_{\rm eq.}}t&\,\,\,t\ll r_h\ll \ell\\
&1&\,\,\,r_h\ll t\ll \ell 
\end{array}.
\eea
It is worth to mention that the value of the constant rate corresponding to the linear growth regime precisely matches with the previous results of 
\cite{Moosa:2020vcs,2001.05501}. Also note that this value is the same as the entanglement velocity found for a BTZ black brane in \cite{Liu:2013qca}. 
\item Higher dimension thermal quench is very similar to the mentioned d=2 case. In this situation, our numerical results, allowing us to identify regimes of early time quadratic growth, an intermediate linear growth and a saturation regime. Moreover, we found that for small entangling regions, EWCS transitions between vacuum and thermal values, without much of a linear regime in between and as the width of the entangling region becomes larger, the region with linear scaling becomes more pronounced. In particular, for a four dimensional bulk theory, in the linear growth regime, we found $\mathcal{R}_W(t)\sim v_w$ where the best fit gives
$v_w\approx 0.68$. Once again, this constant rate precisely matches the entanglement velocity found for a $(3+1)$-dimensional AdS black brane in \cite{Liu:2013qca}. We expect that the same matching happens in higher dimensional cases. It would be interesting
to understand whether this indeed happens to employ an analytic approach similar to three dimensional case.

\item Considering an electromagnetic quench in higher dimensions with $T\neq 0$, we see that, the qualitative features of the evolution is similar to the previous case with $q = 0$. The main conclusion is that turning on additional charged matter fields slow down the thermalization process. On the other hand, choosing the system to be entirely non-thermal by approaching the extremal black brane solution, there exist some interesting differences when comparing the behavior of the evolution to the thermal quench. In this case, the regime of linear growth is replaced by a logarithmic growth. This behavior is inherited from the logarithmic scaling in the static extremal geometries, which has its origin in the double zero at the horizon.

\item 
As we have mentioned before, it is proposed that EWCS is dual to different information
measures including entanglement of purification, reflected entropy, odd entropy and logarithmic
negativity. Considering $E_W$ as a measure of correlations dual to the reflected entropy, figure \ref{fig:EWd3largel} (left panel) indicates that the correlation grows even after $t \sim \mathcal{O}(\frac{\ell}{2})$ where HMI (a natural measure of correlations) reaches its maximum value. It implies that $E_W$ captures more correlations than HMI. This behavior is consistent with the result of \cite{1907.12555,1909.06790,1902.02369} that points out $E_W$  is more sensitive to classical correlations. Although, this interpretation conflicts with the relation between EWCS and negativity (which is just sensitive to quantum correlations) \cite{2001.05501}. Similar behavior for $E_W$  has been noted in \cite{2001.05501} for a different model of the quench in (1+1) dimensional CFT. Here, we emphasize that our study shows this is also true for thermal and electromagnetic quench in (1+1) and (1+2) dimensions. On the other hand, we show the time dependence of (holographic) odd entropy given by eq. \eqref{odd} for various values of $\ell$ in $d=2$ in the right panel of figure \ref{fig:EWd3largel}. The figure demonstrates that this quantity saturates after $t \sim \mathcal{O}(\ell)$ to a constant value which is larger than the vacuum contribution. 
\begin{figure}
	\begin{center}
		\includegraphics[scale=0.75]{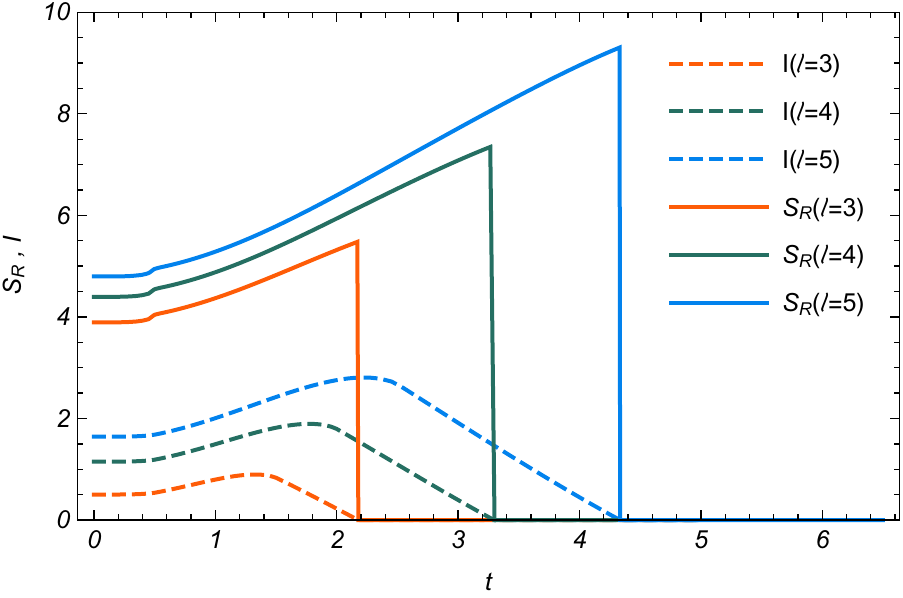}
		\hspace*{1cm}
		\includegraphics[scale=0.75]{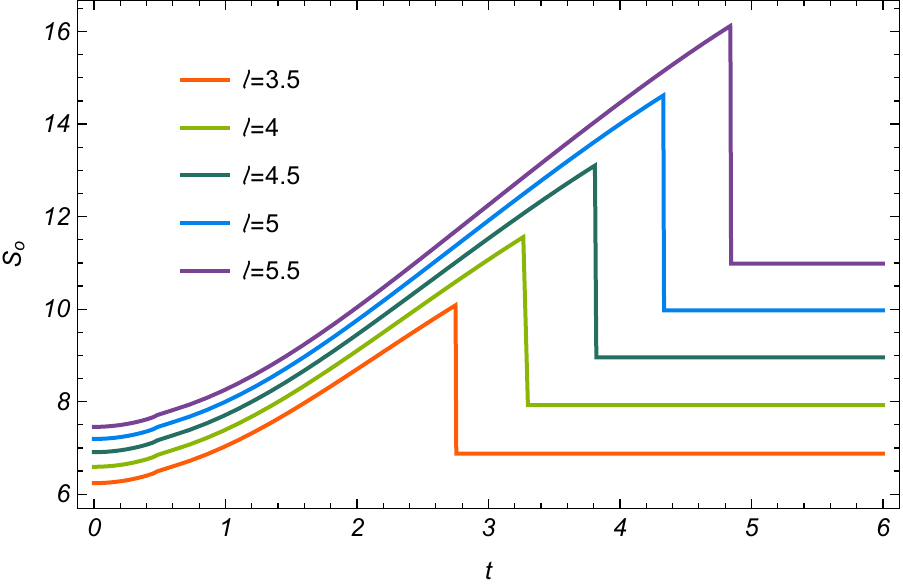}
	\end{center}
	\caption{Evolution of reflected entropy and mutual information (left) and odd entropy (right) in $d=2$. Here we have set $h=1$.}
	\label{fig:EWd3largel}
\end{figure}

\item	EWCS satisfies some inequalities e.g. eqs. \eqref{ineq1} to \eqref{ineq3}. These relations provide important pieces of evidence for finding holographic dual of EWCS \cite{1708.09393,1709.07424,1905.00577}. It is worthwhile to check these inequalities in the time-dependent Vaidya background explicitly. However, in this paper, we restricted our discussion on symmetric configurations where the size of the two subregions is equal \eqref{stripregion} so it only lets us check  eqs.
\eqref{ineq1} and \eqref{ineq2}
in a certain condition.	For example, figure \ref{fig:inequalities}  shows that eqs. \eqref{ineq1} and \eqref{ineq2} hold in the our setup \footnote{One may note that figure \ref{fig:EWd3largel} (left panel) also implies $S_R(A:B)=2E_W(A:B)>I(A:B)$ (see eq. \eqref{ineq1} )}. To check eq. \eqref{ineq3} one should study more general non-symmetric setups. 
\end{itemize}

\begin{figure}
\begin{center}
	\includegraphics[scale=0.75]{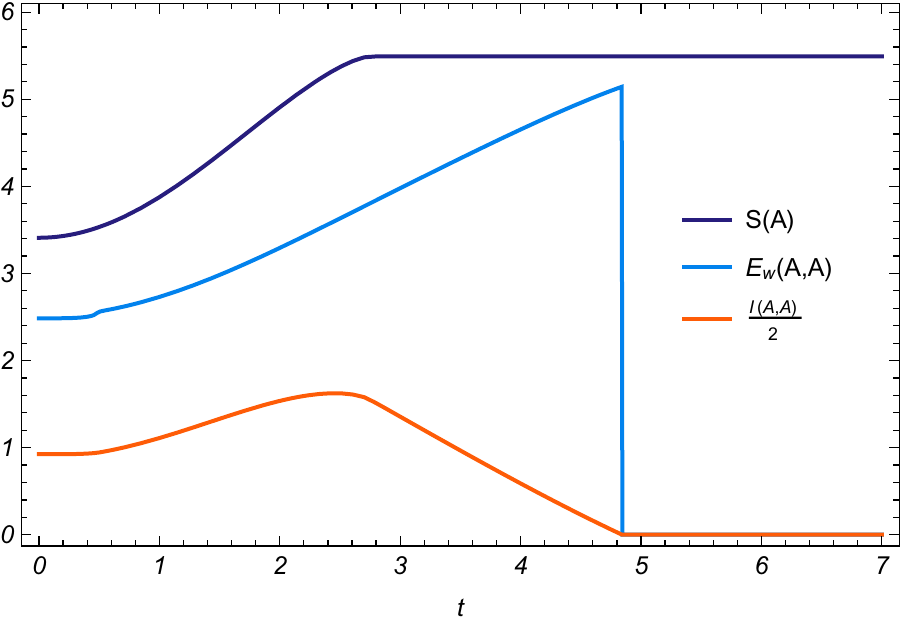}
	\hspace*{1cm}
	\includegraphics[scale=0.75]{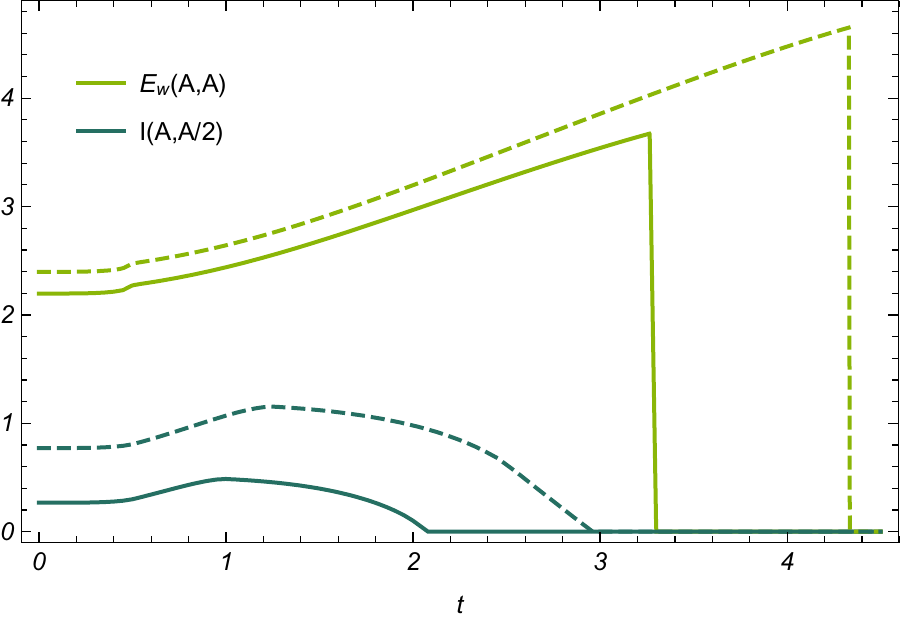}
\end{center}
\caption{\textit{Left}: Plot of the HEE, EWCS and half the HMI as a function of boundary time for $\ell=5.5$. \textit{Right}: Plot of EWCS($\ell$, $\ell$) and HMI($\ell$, $\ell/2$) for $\ell=4$ (solid) and $\ell=5$ (dashed). Here we have set $h=1$ and $d=2$.}
\label{fig:inequalities}
\end{figure}

As we have mentioned before, the symmetric configuration \eqref{stripregion}
significantly simplifies the calculation of the EWCS. In \cite{1902.02243} the authors discuss an efficient algorithm to compute the EWCS in general configurations, which could be applied in the time-dependent case as well. In this case, denoting the widths of the strips as $\ell_1$ and $\ell_2$, the EWCS corresponds to the length of the minimum geodesic connecting $r_d\in \Gamma_{h}$ and $r_u \in \Gamma_{\ell_1+\ell_2+h}$. Note that an important difference is that for non-symmetric configurations, $r_d$ and $r_u$ are no longer coincide with the corresponding turning points of the HRT extremal surfaces, but rather vary as we change the ratio of $\frac{\ell_2}{\ell_1}$. Hence our previous
arguments that the minimum geodesic lies on $x=0$ slice no longer apply,
and we must undertake a more extensive analysis of all possible geodesics ending on the $\Gamma$s. Following \cite{1902.02243}, similar algorithm may be employed to find the minimum geodesic for the Vaidya geometry through an iterative scheme at each time step.  
Even though this would be computationally very demanding, a significant simplification occurs for the case of homogeneous backgrounds, e.g., eq.\eqref{vaidyametric}, in which different geodesics are related by translations along the boundary spatial coordinates.

We can extend this study to different interesting directions. 
Although in higher dimensions we did a numerical analysis, we expect that an analytic treatment similar to the three dimensional case most simply done by considering the thin shell and large entangling region limits. Then one can extract some analytic behavior of EWCS
in different scaling regimes during the evolution which may useful to more investigate interesting features of the thermalization process. In particular it enables us to study various scaling regimes, generalizing the tsunami picture \cite{Liu:2013qca}.

In this paper we restricted our discussion to the equilibration following a global quench in relativistic setup. It is interesting to
consider more general backgrounds, in particular those with Lifshitz and hyperscaling violating exponents \cite{Alishahiha:2014cwa,Fonda:2014ula}. Another interesting direction is to consider small entangling regions similar to \cite{Kundu:2016cgh}. We leave the details of some interesting problems for future study \cite{WORK}.


\subsection*{Acknowledgements}
We are very grateful to Mohsen Alishahiha for careful reading
of the manuscript. We would also like to thank the referee for
her/his useful comments. Mohammad Reza Mohammadi Mozaffar would like to acknowledge the financial support of University of Guilan for this research under grant number 15P-112083.

\end{document}